\documentclass[aps, pra, twocolumn, superscriptaddress, final]{revtex4}
\pdfoutput=1

\usepackage[utf8]{inputenc}
\usepackage[T1]{fontenc}
\usepackage{amsmath}
\usepackage{amssymb}
\usepackage{float}
\usepackage{graphicx}
\usepackage{microtype}
\usepackage{tikz}

\newcommand{\abs}[1]{\left| #1 \right|}
\DeclareMathOperator{\sech}{sech}

\newcommand*\circled[1]{
  \tikz[baseline=(char.base)]{
    \node[shape=circle,draw,inner sep=0.5pt] (char) {#1};
  }
}

\begin{document}

\title{Cherenkov radiation and scattering of external dispersive waves by two-color solitons}

\author{Ivan Oreshnikov}
\email{oreshnikov.ivan@gmail.com}
\affiliation{Max Planck Institute for Intelligent Systems, Max-Planck-Ring 4, 72076 T\"ubingen, Germany}

\author{Oliver Melchert}
\affiliation{Cluster of Exellence PhoenixD, Welfengarten 1, 30167 Hannover, Germany}
\affiliation{Leibniz Universit\"at Hannover, Institute of Quantum Optics, Welfengarten 1, 30167 Hannover, Germany}
\affiliation{Hannover Centre for Optical Technologies, Nienburger Strasse 17, 30167 Hannover, Germany}

\author{Stephanie Willms}
\affiliation{Cluster of Exellence PhoenixD, Welfengarten 1, 30167 Hannover, Germany}
\affiliation{Leibniz Universit\"at Hannover, Institute of Quantum Optics, Welfengarten 1, 30167 Hannover, Germany}

\author{Surajit Bose}
\affiliation{Cluster of Exellence PhoenixD, Welfengarten 1, 30167 Hannover, Germany}
\affiliation{Leibniz Universit\"at Hannover, Institute of Photonics, Nienburger Strasse 17, 30167 Hannover, Germany}

\author{Ihar Babushkin}
\affiliation{Cluster of Exellence PhoenixD, Welfengarten 1, 30167 Hannover, Germany}
\affiliation{Leibniz Universit\"at Hannover, Institute of Quantum Optics, Welfengarten 1, 30167 Hannover, Germany}

\author{Ayhan Demircan}
\affiliation{Cluster of Exellence PhoenixD, Welfengarten 1, 30167 Hannover, Germany}
\affiliation{Leibniz Universit\"at Hannover, Institute of Quantum Optics, Welfengarten 1, 30167 Hannover, Germany}
\affiliation{Hannover Centre for Optical Technologies, Nienburger Strasse 17, 30167 Hannover, Germany}

\author{Uwe Morgner}
\affiliation{Cluster of Exellence PhoenixD, Welfengarten 1, 30167 Hannover, Germany}
\affiliation{Leibniz Universit\"at Hannover, Institute of Quantum Optics, Welfengarten 1, 30167 Hannover, Germany}
\affiliation{Hannover Centre for Optical Technologies, Nienburger Strasse 17, 30167 Hannover, Germany}

\author{Alexey Yulin}
\affiliation{Department of Physics, ITMO University, Kronverskiy pr. 49, 19701 St.~Petersburg, Russia}

\date{\today}

\begin{abstract}
For waveguides with two separate regions of anomalous dispersion, it is
possible to create a quasi-stable two-color solitary wave.
In this paper we consider how those waves interact with dispersive radiation,
both generation of Cherenkov radiation and scattering of incident dispersive
waves.
We derive the analytic resonance conditions and verify them through numeric
experiments. We also report incident radiation driving the internal
oscillations of the soliton during the scattering process in case of an intense
incident radiation.
We generalize the resonance conditions for the case of an oscillating soliton
and demonstrate how one can use the scattering process to probe and excite an
internal mode of two-color soliton molecules.
\end{abstract}

\maketitle

\section{Introduction}

Solitary waves or solitons are localized nonlinear excitations that preserve
their shapes during evolution. In the systems close to integrable ones solitons
are proven to be robust and in many cases different interactions result
only in the variation of the soliton parameters, such as its intensity or
frequency.
In a certain sense solitons can be called ``eigenmodes of the nonlinear
problem'', meaning that the dynamics of the system can be considered as a set
of solitons interacting with quasi-linear dispersive waves.
Due to their robustness, solitons are of fundamental as well as practical
importance, for example, in the context of optical supercontinuum
generation~\cite{RevModPhys.78.1135, RevModPhys.82.1287,
dudley2010supercontinuum} or soliton fiber
lasers~\cite{Mollenauer:84,Kafka:89,Matsas_1992, Turitsyn_2016}.

The mutual interaction of optical solitons in fibers can enable the formation
of bound states, often referred to as soliton molecules.
They can be realized via dispersion engineering in the framework of the
standard nonlinear Schrödinger equation (NSE) and consist of two pulses that
maintain a fixed separation in time \cite{Stratmann:PRL:2005,Hause:PRA:2008}.
Further, optical soliton clusters have been discovered and studied in a large
variety of physical systems described by different equations such as the
generalized nonlinear Schr\"odinger equation~\cite{PhysRevE.51.3572,
PhysRevE.47.2874, PhysRevA.80.043829}, coupled NLSs~\cite{Kivshar:89,
Abdullaev:89}, the Ginzburg-Landau
equation~\cite{PhysRevLett.79.4047, Haelterman:97}, Lugiato-Lefever
equation~\cite{Weng_LugiatoLefever} and many others~\cite{Neshev:03,
Afanasjev:94, PhysRevE.67.056610, Chen:00, PhysRevLett.89.273902,
PhysRevLett.122.084101, PhysRevLett.121.023905, PhysRevLett.118.243901,
PhysRevE.86.036606, Grelu_NatPhoton, TURMANOV20151828}.  The aforementioned
bound states of solitons have a single central frequency and the whole spectrum
is localized around this frequency.

Another possibility to observe soliton molecules is to provide interaction
between solitons having their carrier frequencies well detuned from each
other. To make the interaction efficient the velocities of the solitons must be
close. In the case of scalar solitons this means that we need higher order
dispersion with two different frequency ranges where the solitons can be
accommodated. These solitons were recently reported
in~\cite{PhysRevLett.123.243905, PhysRevA.101.043822, Melchert:21,
Melchert_SciRep21} for pulses propagating in conservative fibers with higher
order dispersion. Similar solitons were discovered also in mode-locked cavity
lasers~\cite{Lourdesamy_NaturePhys} and in coherently pumped ring
resonators~\cite{Moille:18, Melchert:20}.

In the case of nonintegrable systems, solitons can interact with dispersive
waves (DWs) of low intensity and this interaction leads to interesting physical
phenomena, as the efficient generation of new frequencies~\cite{Yulin:04,
PhysRevE.72.016619, Efimov:04, PhysRevLett.95.213902, Efimov:06}.
It should be noted that this effect is closely related to the optical push
broom effect~\cite{deSterke:92}.
Such resonant scattering can be cascaded, enriching optical
supercontinuum spectra significantly~\cite{Gorbach_NatPhoton}.
Similar effects
were studied in Refs.~\cite{PhysRevLett.110.233901,  Demircan_SciRep,
Philbin_Science}, where the term ``optical event horizon'' was coined. It was
also established that resonant dispersive waves affect the parameters of the
solitons and can lead to dispersive wave mediated acceleration of
solitons~\cite{PhysRevE.72.016619, PhysRevLett.106.163901,
PhysRevLett.110.233901, Yulin:13, Demircan:14,  Wang:15, PhysRevA.98.023833}.
It should be mentioned that resonant scattering of dispersive waves is also
studied for dark solitons~\cite{Oreshnikov:15, Marest:18,  Deng:18,
PhysRevA.103.023505}, and oscillating solitons~\cite{Kodama:94,
Conforti_SciRep, Driben:15, PhysRevLett.115.223902, PhysRevLett.116.183901,
PhysRevA.96.013809}.

The present paper aims to study in detail the interaction of dispersive waves
with two-color soliton molecules, consisting of two bound solitons having well separated
frequencies.
In the prior work~\cite{PhysRevLett.123.243905} we have demonstrated how to
create quasi-stable configurations of two tightly-coupled pulses in a
dispersion landscape $\beta(\omega)$ with two regions of anomalous dispersion,
separated by a region of normal dispersion. Each of these subpulses propagates
on its own carrier frequency.
This coupled state, especially in the process of initial
evolution, sheds dispersive waves that resemble Cherenkov radiation that is
observed for other types of solitary waves in a vast variety of
settings~\cite{akhmediev1995cherenkov, afanasjev1996effect, Driben:15,
Conforti_SciRep, PhysRevLett.115.223902, PhysRevA.96.013809}. And since DW
generation is demonstrated, it is reasonable to anticipate that two-color
soliton molecules colliding with DWs will produce resonant emission in the same
way as it occurs in the case of conventional solitons.
Thanks to a more complex structure of the two-color solitons, one can expect
the scattering dynamics to be much richer compared to the conventional single
soliton case. For instance, it is possible that the internal degrees of freedom
can be excited due to the interaction with a DW and this should
strongly affect the dynamics of the system.
Subsequently we confirm this influence, providing a combined analytical and
numerical view of the process.

The paper is structured as follows.
In Sect.~\ref{sec:analytic_model} we introduce a mathematical model for the
light propagating in a nonlinear fiber with higher order dispersion and derive
the condition of resonant four-wave mixing of the soliton molecules with DWs.
In Sect.~\ref{sec:num_exp} we report the results of the numerical simulations
of the different propagation regimes of bichromatic soliton molecules. The
results of the simulations are compared against analytical resonance
conditions.
Section~\ref{sec:nonlin_eff} discusses the weakly nonlinear case where the
intensity of the dispersive waves is sufficiently large to modify the
interaction between the soliton molecules and the waves.  The excitation of the
internal mode is also discussed in this section.
The paper concludes with a summary in Sect.~\ref{sec:conclusions}.

\section{Analytical model\label{sec:analytic_model}}%

In this section we derive a perturbation theory that explains the resonance
conditions that were proposed before in \cite{PhysRevLett.123.243905},
demonstrate additional Cherenkov radiation mechanisms due to four-wave mixing
(FWM) between the frequency components of the soliton, and then extend this
theory to describe the process of scattering of external waves on the soliton.

We start by considering a non-envelope version of a nonlinear Schr\"odinger
equation~\cite{amiranashvili2010hamiltonian}
\begin{equation}
  \label{eq:GNLSE}
  i \partial_{z} \tilde u
    + \beta(\omega) \tilde u
    + \gamma(\omega) \mathcal{F} \left\{
      \abs{u}^{2} u
    \right\}_{(\omega > 0)} = 0.
\end{equation}
Here and further, $\tilde{u}\equiv \tilde{u}(z,\omega)$ indicates the Fourier
image of the field $u\equiv u(z,t)$ and $\mathcal{F}\left\{ \cdots
\right\}_{(\omega > 0)}$ is the explicit Fourier transform taken for the positive
frequencies only.

\begin{widetext}

\noindent To build the perturbation theory let us introduce an ansatz that
represents the solution as a sum of two single-frequency solitary waves $U_{1,
2}(z, t)$ and a small residue radiation $\psi(z, t)$
\begin{gather}
  \label{eq:PerturbationAnsatz}
  u(z, t) = U_{1}(z, t) + U_{2}(z, t) + \psi(z, t), \\
  \abs{\psi} \ll \abs{U_{1}} \sim \abs{U_{2}}. \nonumber
\end{gather}
For the solitary waves $U_{1, 2}(z, t)$ we assume that they satisfy a pair of
coupled nonlinear Schr\"odinger equations below
\begin{equation}
  \label{eq:CoupledSolitons}
  i \partial_{z} U_{n}
    + \beta_{n}(i \partial_{t}) U_{n}(z, t)
    + \gamma(\omega_{n}) \left(
      \abs{U_{n}}^{2} U_{n} + 2 \abs{U_{m}}^{2} U_{n}
    \right) = 0,
\end{equation}
where $n = 1, 2$ and $m = 2, 1 \ne n$. Essentially, this is an assumption that
the two-color soliton molecule consists of two pulses, which are incoherently
coupled by inducing a refractive-index potential on each
other~\cite{agrawal2013nonlinear}.  Each subpulse exists in some
sort of truncated dispersion landscape, defined by the operator $\beta_{n}(i
\partial_{t})$.
Under specific conditions, the subpulses are given by fundamental solitons of a
modified NLS \cite{Melchert:21}.
A reasonable guess for the truncated operator is a parabolic approximation close
to the carrier frequency
\begin{equation}
  \label{eq:TruncatedDispersionOperator}
  \beta_{n}(i \partial_{t}) =
    \beta(\omega_{n}) +
    i \beta'(\omega_{n}) \partial_{t} -
    \frac{1}{2} \, \beta''(\omega_{n}) \partial_{t}^{2}.
\end{equation}
We additionally suppose that in the soliton's frame of reference the envelope evolves with wavenumber $k_{n}(\omega)$ which we approximate by the wavenumber of the fundamental soliton in the ordinary nonlinear Schr\"odinger equation and a correction from a secondary soliton as
\begin{equation}
  \label{eq:SolitonWavenumber}
  k_{n}(\omega)
    \approx \frac{\gamma(\omega_{n}) A_{n}^{2}}{2}
    + \beta(\omega_{n}) + \beta'(\omega_{n}) (\omega - \omega_{n})
    + \gamma(\omega_{n}) A_{m}^{2},
\end{equation}
where $n = 1, 2$, $m = 2, 1 \ne n$ and $A_{n}$ is the soliton amplitude.

By substituting ansatz~\eqref{eq:PerturbationAnsatz} into
Eq.~\eqref{eq:GNLSE}, linearizing with respect to perturbation $\psi(z,
\omega)$, and discarding the terms corresponding to the soliton
Eqs.~\eqref{eq:CoupledSolitons}, we get the equation for $\tilde \psi$
\begin{multline}
  \label{eq:PerturbationEquation}
  i \partial_{z} \tilde \psi
    + \beta(\omega) \tilde \psi
    + \gamma(\omega) \mathcal{F}\left\{
      2 \abs{U_{1} + U_{2}}^{2} \psi +
      \left( U_{1} + U_{2} \right)^{2} \psi^{*}
    \right\}_{(\omega > 0)} = \\
    - \left[ \beta(\omega) - \beta_{1}(\omega) \right] \tilde U_{1}
    - \left[ \beta(\omega) - \beta_{2}(\omega) \right] \tilde U_{2}
    - \gamma(\omega) \mathcal{F} \left\{
      U_{2}^{2} U_{1}^{*} + U_{1}^{2} U_{2}^{*}
    \right\}_{(\omega > 0)}.
\end{multline}
On the right-hand side (RHS) of Eq.~\eqref{eq:PerturbationEquation} we see two
types of driving terms and each of those terms can be in resonance with the
linear DWs that exist in the system if the particular wavenumber
$k(\omega_{*})$ of the driving term is equal to the wavenumber of a DW
$\beta(\omega_{*})$ at some frequency
$\omega_{*}$~\cite{akhmediev1995cherenkov, Yulin:04}. The first type
is given by $\left[ \beta(\omega) - \beta_{n}(\omega) \right] \tilde U_{n}(z,
\omega)$ which drives the generation of Cherenkov radiation by an individual
soliton $U_{n}$ if the following resonance condition is satisfied
\begin{equation}
  \label{eq:CherenkovRadiationResonanceCondition}
  \beta(\omega) = k_{n}(\omega).
\end{equation}
The second type of the terms are $\gamma(\omega) \mathcal{F}\left\{ U_{2}^{2}
U_{1}^{*} \right\}$ and $\gamma(\omega) \mathcal{F}\left\{ U_{1}^{2} U_{2}^{*}
\right\}$ and they correspond to the process of four-wave mixing that in our
case results in the generation of dispersive radiation at some frequency where
\begin{equation}
  \label{eq:FWMRadiationResonanceCondition}
  \beta(\omega) = 2 k_{n}(\omega)  - k_{m}(\omega).
\end{equation}

Let us move on to the problem of external dispersive wave scattering. For that,
we split the perturbation into the incident and the scattered parts
\begin{equation}
  \label{eq:PerturbationSplit}
  \psi(z, t) = \psi_{\text{inc}}(z, t) + \psi_{\text{sc}}(z, t).
\end{equation}
We explicitly define $\psi_{\text{inc}}(z, t)$ as a linear wave that is propagating in a soliton-free medium
\begin{equation}
  \label{eq:IncidentFieldEquation}
  i \partial_{z} \tilde \psi_{\text{inc}}(z, t)
    + \beta(\omega) \tilde \psi_{\text{inc}}(z, t) = 0.
\end{equation}
Substituting Eq.~\eqref{eq:PerturbationSplit}
into~\eqref{eq:PerturbationEquation} and eliminating the terms corresponding to
Eq.~\eqref{eq:IncidentFieldEquation} we are left with the equation for the
scattered component
\begin{multline}
  \label{eq:ScatteredFieldEquation}
  i \partial_{z} \tilde \psi_{\text{sc}}
    + \beta(\omega) \tilde \psi_{\text{sc}}
    + \gamma(\omega) \mathcal{F}\left\{
      2 \abs{U_{1} + U_{2}}^{2} \psi_{\text{sc}} +
      \left( U_{1} + U_{2} \right)^{2} \psi_{\text{sc}}^{*}
    \right\}_{(\omega > 0)} = \text{\ldots (omitted is the RHS of Eq.~\eqref{eq:PerturbationEquation})} \\
    - \gamma(\omega) \mathcal{F}\left\{
      2 \left(
        \abs{U_{1}}^{2} + U_{1} U_{2}^{*} + U_{2} U_{1}^{*} + \abs{U_{2}}^{2}
      \right) \psi_{\text{inc}} +
      \left(
        U_{1}^{2} + 2 \, U_{1} U_{2} + U_{2}^{2}
      \right) \psi_{\text{inc}}^{*}
    \right\}_{(\omega > 0)}.
\end{multline}
\end{widetext}

In addition to the resonance terms already discussed in
Eq.~\eqref{eq:PerturbationEquation}, we see terms that arise due to
interaction between the incident radiation and the soliton. Here, six new
types of resonance behavior are possible. The first one is due to terms
$\abs{U_{1}}^{2} \psi_{\text{inc}}$ and $\abs{U_{2}}^{2} \psi_{\text{inc}}$,
both with the resonance condition
\begin{equation}
  \label{eq:Type0ResonanceCondition}
  \beta(\omega_{\text{sc}}) = \beta(\omega_{\text{inc}}).
\end{equation}
The next two are due to mixed terms $U_{1} U_{2}^{*} \psi_{\text{inc}}$ and $U_{2} U_{1}^{*} \psi_{\text{inc}}$, with the corresponding resonance condition being
\begin{equation}
  \label{eq:Type1ResonanceCondition}
  \beta(\omega_{\text{sc}}) = \pm k_{1} \mp k_{2} + \beta(\omega_{\text{inc}}).
\end{equation}
Another two are due to $U_{n}^{2} \psi_{\text{inc}}^{*}$ and the resonance condition is
\begin{equation}
  \label{eq:Type2ResonanceCondition}
  \beta(\omega_{\text{sc}}) =
    2 k_{n}(\omega_{\text{sc}}) -
    \beta(\omega_{\text{inc}}), ~ n = 1, 2.
\end{equation}
The final one is due to $2 U_{1} U_{2} \psi_{\text{inc}}^{*}$ with the resonances at
\begin{equation}
  \label{eq:Type3ResonanceCondition}
  \beta(\omega_{\text{sc}}) =
    k_{1}(\omega_{\text{sc}}) +
    k_{2}(\omega_{\text{sc}}) -
    \beta(\omega_{\text{inc}}).
\end{equation}
To verify the predictions given by resonance
conditions~\eqref{eq:CherenkovRadiationResonanceCondition},~\eqref{eq:FWMRadiationResonanceCondition},~\eqref{eq:Type0ResonanceCondition},~\eqref{eq:Type1ResonanceCondition},~\eqref{eq:Type2ResonanceCondition},
and~\eqref{eq:Type3ResonanceCondition} we proceed to numeric experiments.

\section{Numerical experiments\label{sec:num_exp}}%

To numerically integrate Eq.~\eqref{eq:GNLSE} we use the integrating factor
method and transform the equation into a non-stiff version for a modified
spectrum~\cite{dudley2010supercontinuum}. The modified equation can be handled
by any standard ODE solver; we use a \texttt{scipy} interface to \texttt{ZVODE}
solver from \texttt{ODEPACK}~\cite{hindmarsh1983odepack, virtanen2020scipy}
(the code necessary to reproduce the results in the paper can be found
in~\cite{sources}). All the computations are performed in a frame of reference
co-moving with the soliton, which is achieved by a transformation
\begin{equation*}
  t \to t - \beta'(\omega_{1}) z,
\end{equation*}
which, in turn, results in
\begin{align*}
  \beta(\omega) &\to
    \beta(\omega)
    - \beta(\omega_{1})
    - \beta'(\omega_{1}) (\omega - \omega_{1}) \\
  k_{1}(\omega) &\to
    \frac{\gamma(\omega_{1}) A_{1}^{2}}{2} + \gamma(\omega_{1}) A_{2}^{2} \\
  k_{2}(\omega) &\to
    \frac{\gamma(\omega_{2}) A_{2}^{2}}{2}
    + \gamma(\omega_{1}) A_{1}^{2}
    + \beta(\omega_{2}) - \beta(\omega_{1}).
\end{align*}
Description of the specific dispersion profile model $\beta(\omega)$ used in the
simulations can be found in the Supplemental Material~\cite{suppMat}.

To study Cherenkov radiation of two color solitons we chain two separate simulations. First, following the prior work~\cite{PhysRevLett.123.243905}, we produce a two-color soliton by integrating an initial condition that is given by a sum of two fundamental solitons of standard NLSE
\begin{equation}
  \label{eq:SeedInitialCondition}
  u_{0}(t) =
      A_{1} \sech(t / T_{1}) e^{-i \omega_{1} t} +
      A_{2} \sech(t / T_{2}) e^{-i \omega_{2} t},
\end{equation}
where frequencies $\omega_{1}$ and $\omega_{2}$ are both lying in the regions of
anomalous dispersion. Frequency $\omega_{1}$ is otherwise arbitrary;
$\omega_{2}$ is chosen so that group velocities of both the frequency components
match
\begin{equation*}
  \beta'(\omega_{1}) = \beta'(\omega_{2}).
\end{equation*}
In most of the simulations presented subsequently, we fix $T_{1} = T_{2} =
20$~fs and the amplitudes $A_{1}$ and $A_{2}$ are chosen as the fundamental
soliton amplitudes at the corresponding frequencies. This configuration sheds a
significant amount of radiation and relaxes to a quasi-stable solitary wave. We
propagate up to $z = 10~\text{cm}$, take the output field of this seed
simulation, and suppress the radiation tails by multiplying it by a
super-Gaussian temporal window, centered on the peak of the soliton molecule.
This isolated soliton molecule serves as an input to the second simulation that is
carried with the same parameters as the original one.

\begin{figure}[t]
  \includegraphics{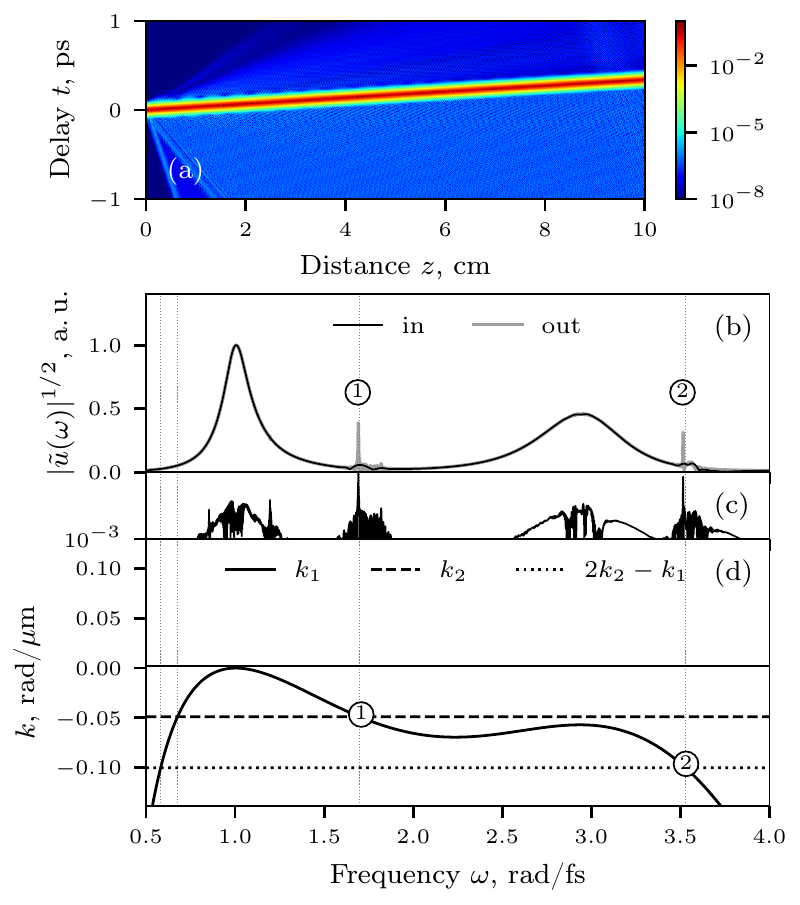}
  \caption{(color online) Cherenkov radiation of an isolated two-color soliton molecule, produced
  by an initial condition with first-component
  carrier frequency $\omega_{1} = 1.010$~rad/fs. (a) Time-domain view, (b)
  input and output spectra, (c) difference between both spectra on a
  logarithmic scale, and (d) diagram of the resonance conditions.}%
  \label{fig:Fig1}
\end{figure}

\begin{figure}[t]
  \includegraphics{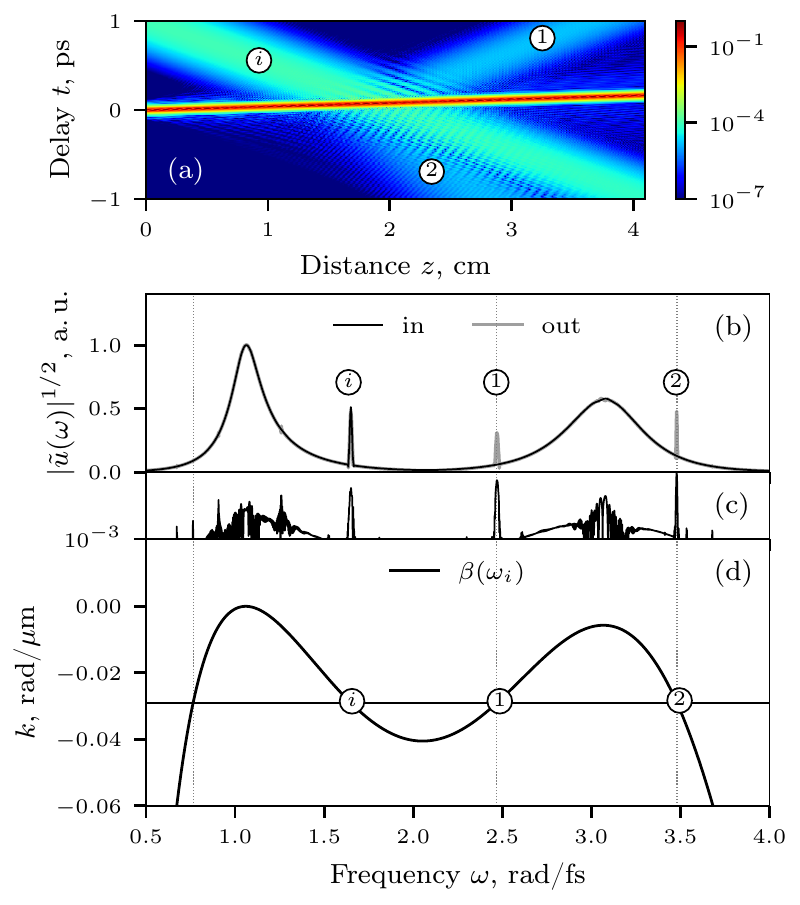}
  \caption{(color online) Scattering of a weak DW with carrier frequency
  $\omega_{\text{inc}} = 1.650$~rad/fs on a two-color soliton with first-component
  carrier frequency $\omega_{1} = 1.070$~rad/fs. (a) Time-domain view, (b)
  input and output spectra, (c) difference between both spectra on a
  logarithmic scale, and (d) diagram of the resonance conditions.}%
  \label{fig:Fig2}
\end{figure}

Once we suppress the radiation that is shed by the seed solitons during the
initial relaxation process the isolated two-color soliton molecule propagates generating only a
narrow-spectrum Cherenkov radiation. An example is shown in
Fig.~\ref{fig:Fig1}. Figure~\ref{fig:Fig1}(b) shows the normalized spectral
densities at the input and output, i.e.\ $z=0\,\mathrm{cm}$ and $10~\mathrm{cm}$, respectively.
For clarity, the difference between both spectra is shown on a logarithmic
scale in Fig.~\ref{fig:Fig1}(c). It is clearly evident that on top of the input
spectrum [thin black line in Fig.~\ref{fig:Fig1}(b)], two additional spectral
lines appear in the output spectrum [gray line in Fig.~\ref{fig:Fig1}(b)].
To clarify the origin of these pronounced spectral lines,
Fig.~\ref{fig:Fig2}(d) demonstrates resonance
conditions~\eqref{eq:CherenkovRadiationResonanceCondition}
and~\eqref{eq:FWMRadiationResonanceCondition}: the black curve corresponds to
$\beta(\omega)$, i.e.\ the left-hand side of both equations;
the horizontal lines correspond to the right-hand sides.
Intersections between the dispersive curve and the horizontal lines that
contribute to Cherenkov radiation are marked separately: \circled{1}~labels
radiation due to the second component of the soliton as predicted by
Eq.~(\ref{eq:CherenkovRadiationResonanceCondition}); \circled{2}~labels the
location of frequencies due to FWM between the frequency components, resulting
in radiation with wavenumber $2 k_{2} - k_{1}$, as predicted
by Eq.~(\ref{eq:FWMRadiationResonanceCondition}).

To study the scattering processes we perform the seed simulation and isolate
the two-color soliton. To accomplish that we add an incident DW in the form of a
Gaussian pulse
\begin{equation*}
  \psi_{\rm{inc}}(t) = A_{\text{inc}} \exp\left(
    - \frac{
      (t - t_{0})^{2}
    }{
      T_{\text{inc}}^{2}
    }
    - i \omega_{\text{inc}} t
  \right).
\end{equation*}
Below we set $A_{\text{inc}}$ to 1\% of the maximum amplitude of the
isolated soliton and fix the width to $T_{\rm{inc}}=300~\mathrm{fs}$.
The initial pulse delay $t_0$ is chosen
as $\pm1000$~fs from the soliton center, with the sign depending on the relative
group velocity between the soliton and the DW and chosen so that
both pulses engage in a collision. Upon propagation
the incident radiation interacts with the two-color soliton to produce
scattered radiation. This process can evolve in several different ways
depending on the frequencies of the soliton and the incident radiation.
Below we consider three concrete examples. In each of them the incident
radiation is split between three different components.

The first configuration, shown in Fig.~\ref{fig:Fig2}, is very close to the
degenerate case where the wavenumbers of the individual soliton components
$k_{1}$ and $k_{2}$ coincide. Equation~\eqref{eq:Type1ResonanceCondition} then
turns into Eq.~\eqref{eq:Type0ResonanceCondition}. This case resembles the
case of fundamental single-component solitons~\cite{Yulin:04}, however, due to the
shape of the dispersive curve $\beta(\omega)$ there is now more than one
nontrivial solution to Eq.~\eqref{eq:Type0ResonanceCondition} and a single
incident frequency yields up to four possible resonances. In practice,
only some of them contribute to the scattered radiation. Those
solutions are separately marked on a diagram on panel~\ref{fig:Fig2}(d):
\circled{$i$} corresponds both to the incident and the partially transmitted
radiation, \circled{1} is the reflected component and \circled{2} is the
additional transmitted component.

\begin{figure}[t]
  \includegraphics{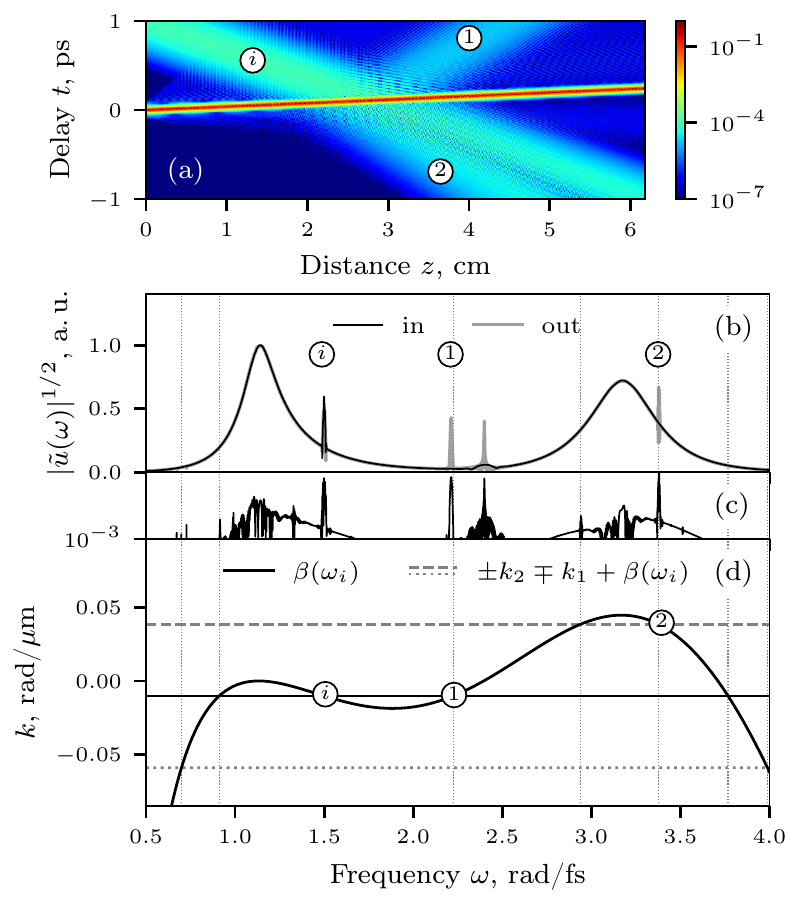}
  \caption{(color online) Scattering of a weak DW with carrier frequency
  $\omega_{\text{inc}} = 1.500$~rad/fs on a two-color soliton with first-component
  carrier frequency $\omega_{1} = 1.150$~rad/fs. (a) Time-domain view, (b)
  input and output spectra, (c) difference between both spectra on a
  logarithmic scale, and (d) diagram of the resonance conditions.}%
  \label{fig:Fig3}
\end{figure}

The second configuration, shown in Fig.~\ref{fig:Fig3}, is a case with
a significant difference between $k_{1}$ and $k_{2}$.
Since Eq.~\eqref{eq:Type1ResonanceCondition} is no longer degenerate, the resonance
diagram in Fig.~\ref{fig:Fig3}(d) is more complex. However, as one can notice, in
that case only the solution marked with \circled{2}, corresponding to the upper
(i.e. ``$-, +$'') branch of Eq.~\eqref{eq:Type1ResonanceCondition}, contributes
to the scattered radiation. Component \circled{1}, corresponding to the
scattered radiation, comes from resonance condition
~\eqref{eq:Type0ResonanceCondition}. The unmarked spectral line close to
$\omega \approx 2.4\,\mathrm{rad/fs}$ is the Cherenkov radiation of the soliton itself.

\begin{figure}[t]
  \includegraphics{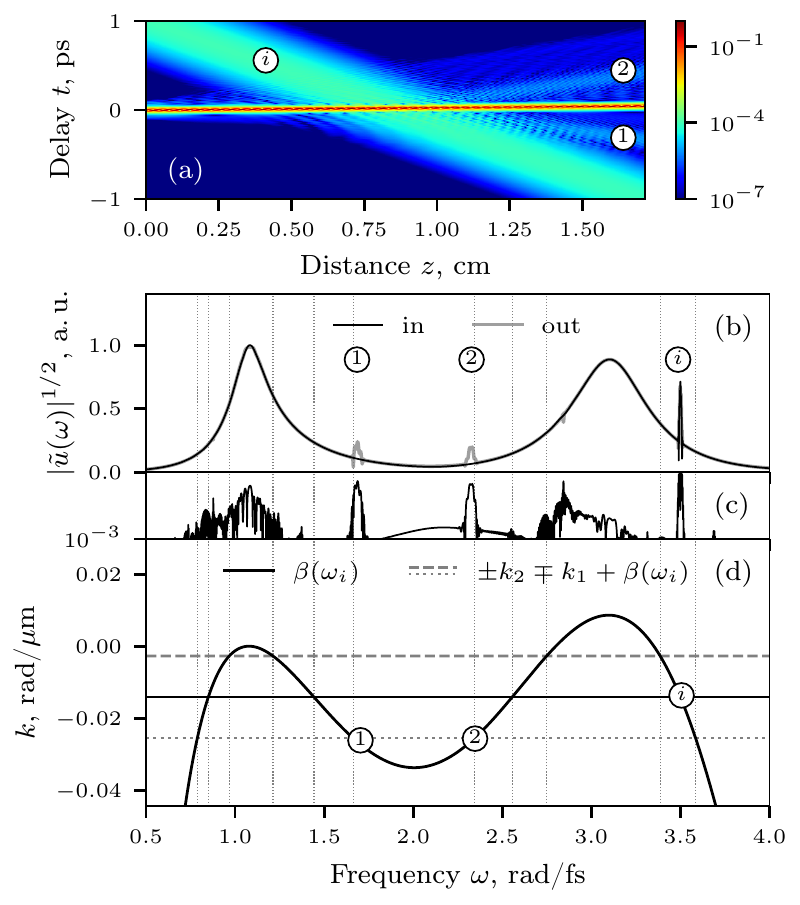}
  \caption{(color online) Scattering of a weak DW with carrier frequency
  $\omega_{\text{inc}} = 3.500$~rad/fs on a two-color soliton with first-component
  carrier frequency $\omega_{1} = 1.150$~rad/fs. (a) Time-domain view, (b)
  input and output spectra, (c) difference between both spectra on a
  logarithmic scale, and (d) diagram of the resonance conditions.}%
  \label{fig:Fig4}
\end{figure}

The third configuration, shown in Fig.~\ref{fig:Fig4}, is in a sense symmetric
to the previous case. Again, the difference between $k_{1}$ and
$k_{2}$ is significant and Eq.~\eqref{eq:Type1ResonanceCondition} is far
from being degenerate, but now the resonances from the lower (i.e.
``$+, -$'') branch of the resonance condition~\eqref{eq:Type1ResonanceCondition} play a significant
role. This is achieved by choosing the incident frequency greater than the
carrier frequency of the second soliton component and enhanced by using an
asymmetric seed soliton with $T_{1} = 30$~fs and $T_{2} = 10$~fs. This creates
a solitary wave with the amplitude of the second component almost equal to the
amplitude of the first one.

Overall, in all of our experiments, only the scattered components corresponding
to Eqs.~\eqref{eq:Type0ResonanceCondition}
and~\eqref{eq:Type1ResonanceCondition} turn out to be significant, while the
components predicted by Eqs.~\eqref{eq:Type2ResonanceCondition}
and~\eqref{eq:Type3ResonanceCondition} do not seem to contribute to the
resulting radiation. In other words, the terms proportional to $\tilde
\psi_{\rm{inc}}^{*}$ on the RHS of
Eq.~\eqref{eq:ScatteredFieldEquation} can be safely neglected.

\section{Nonlinear effects\label{sec:nonlin_eff}}

When deriving resonance
conditions~\eqref{eq:Type0ResonanceCondition}~--~\eqref{eq:Type3ResonanceCondition}
we assumed that the parameters of the solitons themselves stay constant
throughout propagation and scattering processes. This can be considered a
reasonable approximation when the amplitude of the incident wave is negligible
compared to the amplitudes of the individual soliton components. However, in
the general case of more intensive incident radiation this assumption does
not hold. In this section we will briefly discuss two specific examples, where
the process of scattering noticeably affects the parameters of the soliton.
The general setup of the numerical experiments remains as in the preceding
section, i.e.\ we consider scattering of a Gaussian pulse on an isolated
two-color soliton, but this time we increase the amplitude of the DW to
5\% of the soliton's maximum amplitude. This change might seem subtle, but it
is sufficient to make the scattering dynamics much more involved.

\begin{figure}[t]
  \includegraphics{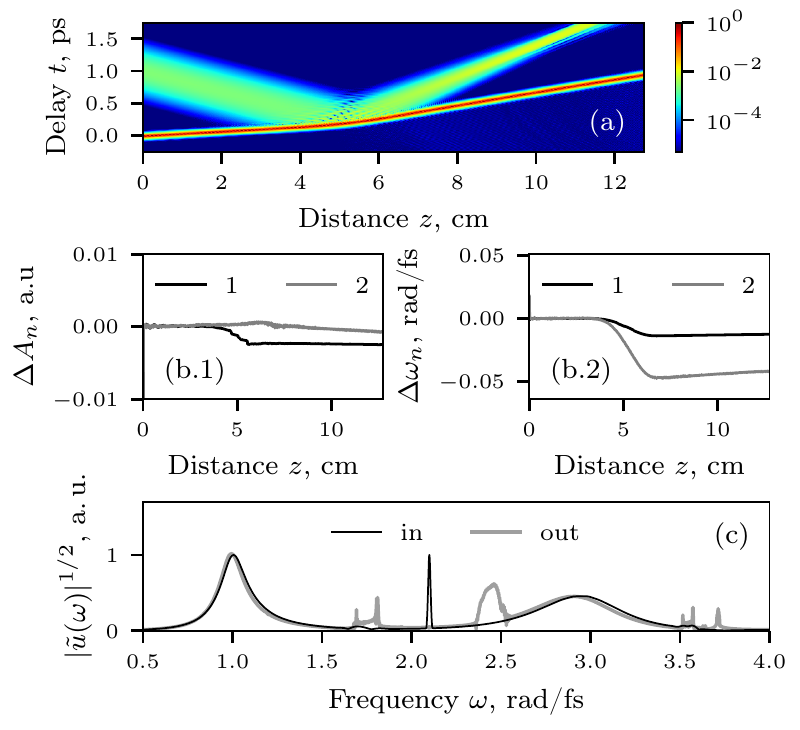}
  \caption{(color online) Scattering of an intensive DW with carrier frequency
  $\omega_{\text{inc}} = 2.100$~rad/fs on a two-color soliton with first-component
  carrier frequency $\omega_{1} = 1.010$~rad/fs. (a) Time-domain view.
  (b.1) Oscillation of the two-color soliton subpulse amplitudes, and, (b.2) oscillation of the
  corresponding center frequencies. (c) Output spectrum.}
  \label{fig:Fig5}
\end{figure}

In the first example, shown in Fig.~\ref{fig:Fig5}, we consider
scattering of a DW with incident frequency $\omega_{i} =
2.100$~rad/fs on a two-color soliton with $\omega_{1} = 1.010$~rad/fs. From
panel~\ref{fig:Fig5}(a) we can immediately notice, that interaction with the DW
significantly decelerates the two-color soliton. This is connected to the change of the
soliton's carrier frequency, as demonstrated in panel~\ref{fig:Fig5}(b.2). This
effect has been demonstrated before for the conventional solitons of nonlinear
Schr\"odinger equation~\cite{PhysRevLett.106.163901, tartara2015soliton}. What
is remarkable about this interaction in case of a two-color soliton is the fact
that the soliton appears to be stable during this process. Granted, the
frequency offset gained by the soliton during the scattering is not especially
prominent [panel~\ref{fig:Fig5}(b.2)], but at the same time the amplitude
difference in both the frequency components stays under 1\%
[panel~\ref{fig:Fig5}(b.1)], so almost no power loss occurs.

\begin{figure}[t]
  \includegraphics{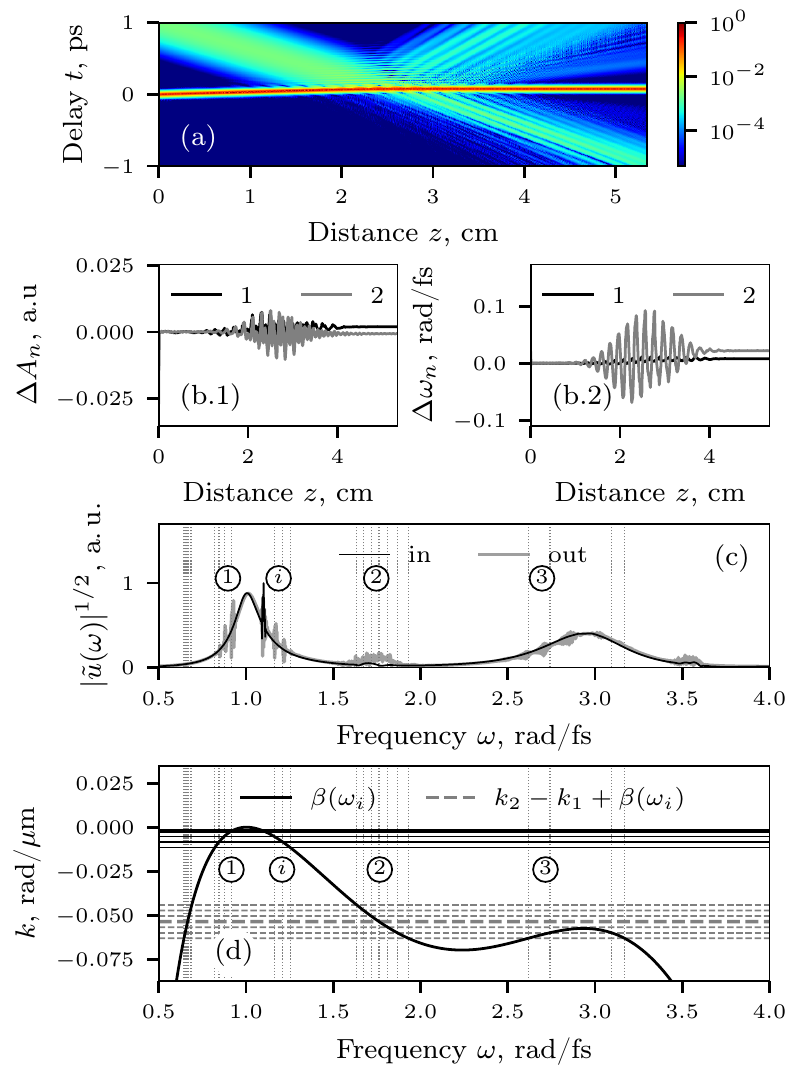}
  \caption{(color online) Scattering of an intensive DW with carrier frequency
  $\omega_{\text{inc}} = 1.100$~rad/fs on a two-color soliton with first-component
  carrier frequency $\omega_{1} = 1.010$~rad/fs. (a) Time-domain view.
  (b.1) Oscillation of the two-color soliton subpulse amplitudes, and, (b.2) oscillation of the
  corresponding center frequencies. (c) Output spectrum. (d) Diagram of the
  resonance conditions.}%
  \label{fig:Fig6}
\end{figure}

In the second example, shown in Fig.~\ref{fig:Fig6}, we consider the
scattering of a DW with the incident frequency $\omega_{i} =
1.100$~rad/fs on a two-color soliton with $\omega_{1} = 1.010$~rad/fs. In the plots in
panels~\ref{fig:Fig6}(b.1) and~\ref{fig:Fig6}(b.2) one can notice how the
interaction with the DW generates oscillations in the soliton parameters; this
is especially prominent in~\ref{fig:Fig6}(b.2) that displays the absolute change
of the soliton's central frequencies $\omega_{1}$ and $\omega_{2}$. From the
latter plot the period of those oscillations can be estimated as $Z_0 =
2~\text{mm}$. The interaction of freely oscillating nonlinear waves, both the
radiation and the scattering processes, is a well studied
problem~\cite{Driben:15, oreshnikov2015interaction,
PhysRevLett.115.223902,Conforti_SciRep,PhysRevA.96.013809} and the common trait
in this setting, independent of the nature of the soliton oscillations, is that
the dispersive radiation produced (generated or scattered) by the soliton is
polychromatic, i.e.\ it consists of several isolated spectral components. This
is indeed what we see in the output spectrum on panel~\ref{fig:Fig6}(c).

To explain this behavior let us return to
equations~\eqref{eq:PerturbationEquation}
and~\eqref{eq:ScatteredFieldEquation}. In the oscillating case the individual
solitons $U_{1}$ and $U_{2}$ are no longer represented by a single spatial
frequency $\propto e^{i k_{1} z}$ and $\propto e^{i k_{2} z}$ and they rather
correspond to Fourier series'
\begin{equation*}
  U_{n}(z, t) =
    \sum \limits_{N \in \mathbb{Z}}
    C_{nN}(t) \exp \left(
      i k_{1} z + i \frac{2 \pi N}{Z_{0}} z
    \right),
\end{equation*}
where $Z_{0}$ is the oscillation period. This leads to a split in resonance conditions, and for the oscillating case equations \eqref{eq:CherenkovRadiationResonanceCondition}, \eqref{eq:FWMRadiationResonanceCondition}, \eqref{eq:Type0ResonanceCondition} and \eqref{eq:Type1ResonanceCondition} read
\begin{gather}
  \label{eq:CherenkovRadiationresonanceConditionZ0}
  \tag{\ref{eq:CherenkovRadiationResonanceCondition}*}
  \beta(\omega) = k_{n}(\omega) + \frac{2 \pi N}{Z_{0}} \\
  \label{eq:FWMradiationResonanceConditionZ0}
  \tag{\ref{eq:FWMRadiationResonanceCondition}*}
  \beta(\omega) =
    2 k_{n}(\omega)  - k_{m}(\omega) +
    \frac{2 \pi N}{Z_{0}} \\
  \label{eq:Type0ResonanceConditionZ0}
  \tag{\ref{eq:Type0ResonanceCondition}*}
  \beta(\omega_{\text{sc}}) =
    \beta(\omega_{\text{inc}}) +
    \frac{2 \pi N}{Z_{0}} \\
  \label{eq:Type1ResonanceConditionZ0}
  \tag{\ref{eq:Type1ResonanceCondition}*}
  \beta(\omega_{\text{sc}}) =
    \pm k_{1} \mp k_{2} +
    \beta(\omega_{\text{inc}}) +
    \frac{2 \pi N}{Z_{0}},
\end{gather}
with $N \in \mathbb{Z}$ being the spatial harmonic number.

The resonance conditions for the scattering process in the last simulation are
displayed on panel~\ref{fig:Fig6}(d). In this process only the last two
equations, i.e.\
Eqs.~(\ref{eq:Type0ResonanceCondition}*) and (\ref{eq:Type1ResonanceCondition}*) are
relevant. The solid black lines correspond to
equation~\eqref{eq:Type0ResonanceCondition} with harmonics $N =
-3, \dots, 0$. This harmonic split not only leads to the reflected part of the
radiation \circled{1} becoming polychromatic, but also has the same effect on
the incident component \circled{i}. Resonance
condition~\eqref{eq:Type1ResonanceCondition} is represented by the gray dashed
lines for harmonics $N = -3, \dots, +3$. It contributes to the scattered
radiation twice: with a wide band \circled{2} that otherwise would degenerate
into a sharp spectral line in a non-oscillating case, and with three wide
spectral lines in the vicinity of \circled{3} which is the contribution of the
lower harmonics $N = -3, -2, -1$. One can notice that there are only two
vertical lines corresponding to the numeric solution
of Eq.~\eqref{eq:Type1ResonanceCondition} displayed around \circled{3}. This is
merely due to a small numeric error in estimating the soliton wavenumber
$k_{2}$ in the plotting procedure, which causes harmonic $N=-1$ to touch the
dispersive rather than intersect it.

\begin{figure}[t]
  \includegraphics{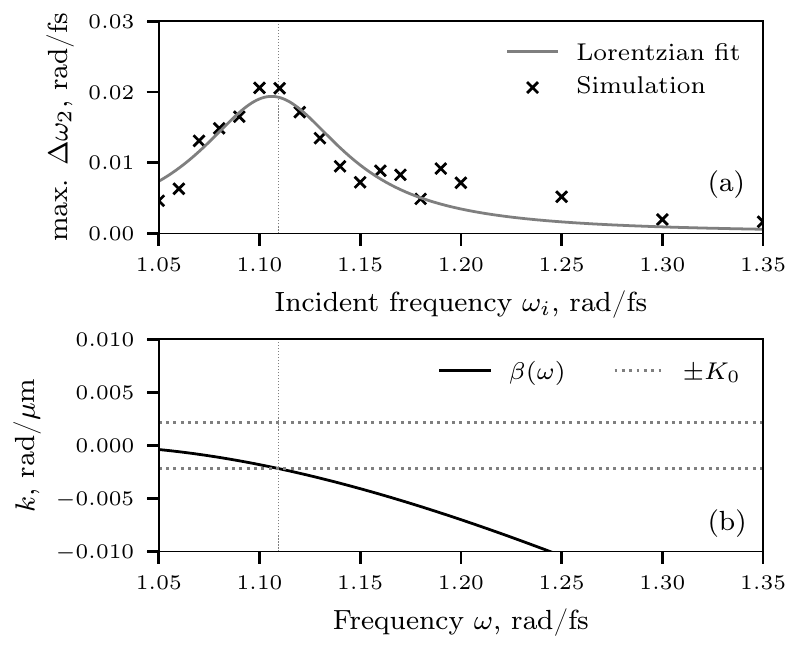}
  \caption{Resonance behavior of the internal mode of the two-color soliton molecule
  during intensive scattering.
  (a) Oscillation of the center frequency of the second component of the two-color soliton
  as function of the frequency of the incident DW.
  (b) Diagram of the resonance condition for the internal mode.
  \label{fig:Fig7}}%
\end{figure}

The specific oscillation mode we see in Fig.~\ref{fig:Fig6} appears to be
heavily damped, since launching the initial soliton with an additional
frequency detuning does not lead to free frequency oscillations during
the propagation.
Such internal dynamics, reminiscent of molecular vibrations, were reported also
previously \cite{PhysRevLett.123.243905,Melchert_SciRep21}, and the emission of
DWs by the two-color soliton was found to explain the dampening of the
oscillation mode \cite{Willms:PRA:2022}.
Here, the incident radiation drives this oscillation, and one could expect to
observe a resonance behavior with respect to the frequency of the incident DW.
And indeed, as demonstrated by the parameter sweep in Fig.~\ref{fig:Fig7}(a),
by adjusting the incident frequency $\omega_{i}$ one can affect, to a certain
degree, the amplitude of the frequency oscillations of the second component
$\Delta \omega_{2}$. For this specific mode, where the frequency oscillations
dominate, it is straightforward to get an estimate for the period $Z_0$ of the
mode and the corresponding wavenumber $K_0$ in terms of a variational approach
(see Supplemental Material \cite{suppMat}).
As shown in Fig.~\ref{fig:Fig7}(b), superimposing the resulting resonance
wavenumber $K_0$ on the dispersion curve allows to graphically estimate the
frequency of the incident DW that corresponds to the internal oscillation mode.
The vertical dashed lines in Figs.~\ref{fig:Fig7}(a,b) indicate this resonance
frequency, confirming the excellent agreement of the numerical simulations and
the approximate analytic estimate.

\section{Conclusion\label{sec:conclusions}}

In this paper, we considered the resonant interaction of two-color soliton
molecules with DWs. The resonance condition for Cherenkov radiation is derived
and analyzed. The comparison with the results of numerical simulations shows
that these conditions correctly predict the positions of the Cherenkov
radiation. We also study the process of the collision of the two-color solitons
with the incident DWs. The resonance conditions predict well all newly
generated frequencies observed in the numerical experiments. Some of the
predicted resonances are not seen in the numerically calculated radiation
spectra. This can be explained by the low efficiency of the corresponding
scattering processes. The theory developed for a simpler case in
\cite{PhysRevE.72.016619} shows that indeed for some parameters the generated
radiation can be extremely weak. The theory giving the analytical estimations
for four waves mixing between the DWs and two-color solitons is of
interest but is out of the scope of the present paper. We also studied how the
intensity of the incident dispersive waves affects the scattering. In
particular, it is shown that the scattering can affect the soliton trajectory
without affecting its integrity. Another important finding is that in the
nonlinear case the DWs can resonantly excite internal oscillations
of the soliton. This results in polychromatic emission of DWs by
oscillating two-color solitons.

\subsection*{Acknowledgements}
OM, SW, IB, AD, and UM acknowledge financial support from Deutsche
Forschungsgemeinschaft (DFG) under Germany’s Excellence Strategy within the
Cluster of Excellence PhoenixD (Photonics, Optics, and Engineering—Innovation
Across Disciplines) (EXC 2122, Project No.\ 390833453).
SB acknowledges financial support from DFG under Germany's Excellence
Strategy within the Cluster of Excellence QuantumFrontiers (EXC 2123, Project
No.\ 390837967).
IB also acknowledges support from DFG (Project No.\ BA4156/4-2).
UM also acknowledges support from DFG (Project No.\ MO 850-20/1).

\bibliographystyle{apsrev4-1}
\bibliography{Bibliography}

\begin{thebibliography}{71}%
\makeatletter
\providecommand \@ifxundefined [1]{%
 \@ifx{#1\undefined}
}%
\providecommand \@ifnum [1]{%
 \ifnum #1\expandafter \@firstoftwo
 \else \expandafter \@secondoftwo
 \fi
}%
\providecommand \@ifx [1]{%
 \ifx #1\expandafter \@firstoftwo
 \else \expandafter \@secondoftwo
 \fi
}%
\providecommand \natexlab [1]{#1}%
\providecommand \enquote  [1]{``#1''}%
\providecommand \bibnamefont  [1]{#1}%
\providecommand \bibfnamefont [1]{#1}%
\providecommand \citenamefont [1]{#1}%
\providecommand \href@noop [0]{\@secondoftwo}%
\providecommand \href [0]{\begingroup \@sanitize@url \@href}%
\providecommand \@href[1]{\@@startlink{#1}\@@href}%
\providecommand \@@href[1]{\endgroup#1\@@endlink}%
\providecommand \@sanitize@url [0]{\catcode `\\12\catcode `\$12\catcode
  `\&12\catcode `\#12\catcode `\^12\catcode `\_12\catcode `\%12\relax}%
\providecommand \@@startlink[1]{}%
\providecommand \@@endlink[0]{}%
\providecommand \url  [0]{\begingroup\@sanitize@url \@url }%
\providecommand \@url [1]{\endgroup\@href {#1}{\urlprefix }}%
\providecommand \urlprefix  [0]{URL }%
\providecommand \Eprint [0]{\href }%
\providecommand \doibase [0]{http://dx.doi.org/}%
\providecommand \selectlanguage [0]{\@gobble}%
\providecommand \bibinfo  [0]{\@secondoftwo}%
\providecommand \bibfield  [0]{\@secondoftwo}%
\providecommand \translation [1]{[#1]}%
\providecommand \BibitemOpen [0]{}%
\providecommand \bibitemStop [0]{}%
\providecommand \bibitemNoStop [0]{.\EOS\space}%
\providecommand \EOS [0]{\spacefactor3000\relax}%
\providecommand \BibitemShut  [1]{\csname bibitem#1\endcsname}%
\let\auto@bib@innerbib\@empty
\bibitem [{\citenamefont {Dudley}\ \emph {et~al.}(2006)\citenamefont {Dudley},
  \citenamefont {Genty},\ and\ \citenamefont {Coen}}]{RevModPhys.78.1135}%
  \BibitemOpen
  \bibfield  {author} {\bibinfo {author} {\bibfnamefont {J.~M.}\ \bibnamefont
  {Dudley}}, \bibinfo {author} {\bibfnamefont {G.}~\bibnamefont {Genty}}, \
  and\ \bibinfo {author} {\bibfnamefont {S.}~\bibnamefont {Coen}},\ }\href@noop
  {} {\bibfield  {journal} {\bibinfo  {journal} {Rev. Mod. Phys.}\ }\textbf
  {\bibinfo {volume} {78}},\ \bibinfo {pages} {1135} (\bibinfo {year}
  {2006})}\BibitemShut {NoStop}%
\bibitem [{\citenamefont {Skryabin}\ and\ \citenamefont
  {Gorbach}(2010)}]{RevModPhys.82.1287}%
  \BibitemOpen
  \bibfield  {author} {\bibinfo {author} {\bibfnamefont {D.~V.}\ \bibnamefont
  {Skryabin}}\ and\ \bibinfo {author} {\bibfnamefont {A.~V.}\ \bibnamefont
  {Gorbach}},\ }\href@noop {} {\bibfield  {journal} {\bibinfo  {journal} {Rev.
  Mod. Phys.}\ }\textbf {\bibinfo {volume} {82}},\ \bibinfo {pages} {1287}
  (\bibinfo {year} {2010})}\BibitemShut {NoStop}%
\bibitem [{\citenamefont {Dudley}\ and\ \citenamefont
  {Taylor}(2010)}]{dudley2010supercontinuum}%
  \BibitemOpen
  \bibfield  {author} {\bibinfo {author} {\bibfnamefont {J.~M.}\ \bibnamefont
  {Dudley}}\ and\ \bibinfo {author} {\bibfnamefont {J.~R.}\ \bibnamefont
  {Taylor}},\ }\href@noop {} {\emph {\bibinfo {title} {Supercontinuum
  generation in optical fibers}}}\ (\bibinfo  {publisher} {Cambridge University
  Press},\ \bibinfo {year} {2010})\BibitemShut {NoStop}%
\bibitem [{\citenamefont {Mollenauer}\ and\ \citenamefont
  {Stolen}(1984)}]{Mollenauer:84}%
  \BibitemOpen
  \bibfield  {author} {\bibinfo {author} {\bibfnamefont {L.~F.}\ \bibnamefont
  {Mollenauer}}\ and\ \bibinfo {author} {\bibfnamefont {R.~H.}\ \bibnamefont
  {Stolen}},\ }\href@noop {} {\bibfield  {journal} {\bibinfo  {journal} {Opt.
  Lett.}\ }\textbf {\bibinfo {volume} {9}},\ \bibinfo {pages} {13} (\bibinfo
  {year} {1984})}\BibitemShut {NoStop}%
\bibitem [{\citenamefont {Kafka}\ \emph {et~al.}(1989)\citenamefont {Kafka},
  \citenamefont {Baer},\ and\ \citenamefont {Hall}}]{Kafka:89}%
  \BibitemOpen
  \bibfield  {author} {\bibinfo {author} {\bibfnamefont {J.~D.}\ \bibnamefont
  {Kafka}}, \bibinfo {author} {\bibfnamefont {T.}~\bibnamefont {Baer}}, \ and\
  \bibinfo {author} {\bibfnamefont {D.~W.}\ \bibnamefont {Hall}},\ }\href@noop
  {} {\bibfield  {journal} {\bibinfo  {journal} {Opt. Lett.}\ }\textbf
  {\bibinfo {volume} {14}},\ \bibinfo {pages} {1269} (\bibinfo {year}
  {1989})}\BibitemShut {NoStop}%
\bibitem [{\citenamefont {Matsas}\ \emph {et~al.}(1992)\citenamefont {Matsas},
  \citenamefont {Newson}, \citenamefont {Richardson},\ and\ \citenamefont
  {Payne}}]{Matsas_1992}%
  \BibitemOpen
  \bibfield  {author} {\bibinfo {author} {\bibfnamefont {V.}~\bibnamefont
  {Matsas}}, \bibinfo {author} {\bibfnamefont {T.}~\bibnamefont {Newson}},
  \bibinfo {author} {\bibfnamefont {D.}~\bibnamefont {Richardson}}, \ and\
  \bibinfo {author} {\bibfnamefont {D.}~\bibnamefont {Payne}},\ }\href@noop {}
  {\bibfield  {journal} {\bibinfo  {journal} {Electron. Lett.}\ }\textbf
  {\bibinfo {volume} {28}},\ \bibinfo {pages} {1391} (\bibinfo {year}
  {1992})}\BibitemShut {NoStop}%
\bibitem [{\citenamefont {Turitsyn}\ \emph {et~al.}(2016)\citenamefont
  {Turitsyn}, \citenamefont {Rosanov}, \citenamefont {Yarutkina}, \citenamefont
  {Bednyakova}, \citenamefont {Fedorov}, \citenamefont {Shtyrina},\ and\
  \citenamefont {Fedoruk}}]{Turitsyn_2016}%
  \BibitemOpen
  \bibfield  {author} {\bibinfo {author} {\bibfnamefont {S.~K.}\ \bibnamefont
  {Turitsyn}}, \bibinfo {author} {\bibfnamefont {N.~N.}\ \bibnamefont
  {Rosanov}}, \bibinfo {author} {\bibfnamefont {I.~A.}\ \bibnamefont
  {Yarutkina}}, \bibinfo {author} {\bibfnamefont {A.~E.}\ \bibnamefont
  {Bednyakova}}, \bibinfo {author} {\bibfnamefont {S.~V.}\ \bibnamefont
  {Fedorov}}, \bibinfo {author} {\bibfnamefont {O.~V.}\ \bibnamefont
  {Shtyrina}}, \ and\ \bibinfo {author} {\bibfnamefont {M.~P.}\ \bibnamefont
  {Fedoruk}},\ }\href@noop {} {\bibfield  {journal} {\bibinfo  {journal}
  {Phys.-Usp.}\ }\textbf {\bibinfo {volume} {59}},\ \bibinfo {pages} {642}
  (\bibinfo {year} {2016})}\BibitemShut {NoStop}%
\bibitem [{\citenamefont {Stratmann}\ \emph {et~al.}(2005)\citenamefont
  {Stratmann}, \citenamefont {Pagel},\ and\ \citenamefont
  {Mitschke}}]{Stratmann:PRL:2005}%
  \BibitemOpen
  \bibfield  {author} {\bibinfo {author} {\bibfnamefont {M.}~\bibnamefont
  {Stratmann}}, \bibinfo {author} {\bibfnamefont {T.}~\bibnamefont {Pagel}}, \
  and\ \bibinfo {author} {\bibfnamefont {F.}~\bibnamefont {Mitschke}},\
  }\href@noop {} {\bibfield  {journal} {\bibinfo  {journal} {Phys. Rev. Lett.}\
  }\textbf {\bibinfo {volume} {95}},\ \bibinfo {pages} {143902} (\bibinfo
  {year} {2005})}\BibitemShut {NoStop}%
\bibitem [{\citenamefont {Hause}\ \emph {et~al.}(2008)\citenamefont {Hause},
  \citenamefont {Hartwig}, \citenamefont {B\"ohm},\ and\ \citenamefont
  {Mitschke}}]{Hause:PRA:2008}%
  \BibitemOpen
  \bibfield  {author} {\bibinfo {author} {\bibfnamefont {A.}~\bibnamefont
  {Hause}}, \bibinfo {author} {\bibfnamefont {H.}~\bibnamefont {Hartwig}},
  \bibinfo {author} {\bibfnamefont {M.}~\bibnamefont {B\"ohm}}, \ and\ \bibinfo
  {author} {\bibfnamefont {F.}~\bibnamefont {Mitschke}},\ }\href@noop {}
  {\bibfield  {journal} {\bibinfo  {journal} {Phys. Rev. A}\ }\textbf {\bibinfo
  {volume} {78}},\ \bibinfo {pages} {063817} (\bibinfo {year}
  {2008})}\BibitemShut {NoStop}%
\bibitem [{\citenamefont {Buryak}\ and\ \citenamefont
  {Akhmediev}(1995)}]{PhysRevE.51.3572}%
  \BibitemOpen
  \bibfield  {author} {\bibinfo {author} {\bibfnamefont {A.~V.}\ \bibnamefont
  {Buryak}}\ and\ \bibinfo {author} {\bibfnamefont {N.~N.}\ \bibnamefont
  {Akhmediev}},\ }\href@noop {} {\bibfield  {journal} {\bibinfo  {journal}
  {Phys. Rev. E}\ }\textbf {\bibinfo {volume} {51}},\ \bibinfo {pages} {3572}
  (\bibinfo {year} {1995})}\BibitemShut {NoStop}%
\bibitem [{\citenamefont {Malomed}(1993)}]{PhysRevE.47.2874}%
  \BibitemOpen
  \bibfield  {author} {\bibinfo {author} {\bibfnamefont {B.~A.}\ \bibnamefont
  {Malomed}},\ }\href@noop {} {\bibfield  {journal} {\bibinfo  {journal} {Phys.
  Rev. E}\ }\textbf {\bibinfo {volume} {47}},\ \bibinfo {pages} {2874}
  (\bibinfo {year} {1993})}\BibitemShut {NoStop}%
\bibitem [{\citenamefont {Zavyalov}\ \emph {et~al.}(2009)\citenamefont
  {Zavyalov}, \citenamefont {Iliew}, \citenamefont {Egorov},\ and\
  \citenamefont {Lederer}}]{PhysRevA.80.043829}%
  \BibitemOpen
  \bibfield  {author} {\bibinfo {author} {\bibfnamefont {A.}~\bibnamefont
  {Zavyalov}}, \bibinfo {author} {\bibfnamefont {R.}~\bibnamefont {Iliew}},
  \bibinfo {author} {\bibfnamefont {O.}~\bibnamefont {Egorov}}, \ and\ \bibinfo
  {author} {\bibfnamefont {F.}~\bibnamefont {Lederer}},\ }\href@noop {}
  {\bibfield  {journal} {\bibinfo  {journal} {Phys. Rev. A}\ }\textbf {\bibinfo
  {volume} {80}},\ \bibinfo {pages} {043829} (\bibinfo {year}
  {2009})}\BibitemShut {NoStop}%
\bibitem [{\citenamefont {Kivshar}\ and\ \citenamefont
  {Malomed}(1989)}]{Kivshar:89}%
  \BibitemOpen
  \bibfield  {author} {\bibinfo {author} {\bibfnamefont {Y.~S.}\ \bibnamefont
  {Kivshar}}\ and\ \bibinfo {author} {\bibfnamefont {B.~A.}\ \bibnamefont
  {Malomed}},\ }\href@noop {} {\bibfield  {journal} {\bibinfo  {journal} {Opt.
  Lett.}\ }\textbf {\bibinfo {volume} {14}},\ \bibinfo {pages} {1365} (\bibinfo
  {year} {1989})}\BibitemShut {NoStop}%
\bibitem [{\citenamefont {Abdullaev}\ \emph {et~al.}(1989)\citenamefont
  {Abdullaev}, \citenamefont {Abrarov},\ and\ \citenamefont
  {Darmanyan}}]{Abdullaev:89}%
  \BibitemOpen
  \bibfield  {author} {\bibinfo {author} {\bibfnamefont {F.~K.}\ \bibnamefont
  {Abdullaev}}, \bibinfo {author} {\bibfnamefont {R.~M.}\ \bibnamefont
  {Abrarov}}, \ and\ \bibinfo {author} {\bibfnamefont {S.~A.}\ \bibnamefont
  {Darmanyan}},\ }\href@noop {} {\bibfield  {journal} {\bibinfo  {journal}
  {Opt. Lett.}\ }\textbf {\bibinfo {volume} {14}},\ \bibinfo {pages} {131}
  (\bibinfo {year} {1989})}\BibitemShut {NoStop}%
\bibitem [{\citenamefont {Akhmediev}\ \emph {et~al.}(1997)\citenamefont
  {Akhmediev}, \citenamefont {Ankiewicz},\ and\ \citenamefont
  {Soto-Crespo}}]{PhysRevLett.79.4047}%
  \BibitemOpen
  \bibfield  {author} {\bibinfo {author} {\bibfnamefont {N.~N.}\ \bibnamefont
  {Akhmediev}}, \bibinfo {author} {\bibfnamefont {A.}~\bibnamefont
  {Ankiewicz}}, \ and\ \bibinfo {author} {\bibfnamefont {J.~M.}\ \bibnamefont
  {Soto-Crespo}},\ }\href@noop {} {\bibfield  {journal} {\bibinfo  {journal}
  {Phys. Rev. Lett.}\ }\textbf {\bibinfo {volume} {79}},\ \bibinfo {pages}
  {4047} (\bibinfo {year} {1997})}\BibitemShut {NoStop}%
\bibitem [{\citenamefont {Haelterman}\ \emph {et~al.}(1997)\citenamefont
  {Haelterman}, \citenamefont {Trillo},\ and\ \citenamefont
  {Ferro}}]{Haelterman:97}%
  \BibitemOpen
  \bibfield  {author} {\bibinfo {author} {\bibfnamefont {M.}~\bibnamefont
  {Haelterman}}, \bibinfo {author} {\bibfnamefont {S.}~\bibnamefont {Trillo}},
  \ and\ \bibinfo {author} {\bibfnamefont {P.}~\bibnamefont {Ferro}},\
  }\href@noop {} {\bibfield  {journal} {\bibinfo  {journal} {Opt. Lett.}\
  }\textbf {\bibinfo {volume} {22}},\ \bibinfo {pages} {84} (\bibinfo {year}
  {1997})}\BibitemShut {NoStop}%
\bibitem [{\citenamefont {Weng}\ \emph {et~al.}(2020)\citenamefont {Weng},
  \citenamefont {Bouchand}, \citenamefont {Lucas}, \citenamefont {Obrzud},
  \citenamefont {Herr},\ and\ \citenamefont
  {Kippenberg}}]{Weng_LugiatoLefever}%
  \BibitemOpen
  \bibfield  {author} {\bibinfo {author} {\bibfnamefont {W.}~\bibnamefont
  {Weng}}, \bibinfo {author} {\bibfnamefont {R.}~\bibnamefont {Bouchand}},
  \bibinfo {author} {\bibfnamefont {E.}~\bibnamefont {Lucas}}, \bibinfo
  {author} {\bibfnamefont {E.}~\bibnamefont {Obrzud}}, \bibinfo {author}
  {\bibfnamefont {T.}~\bibnamefont {Herr}}, \ and\ \bibinfo {author}
  {\bibfnamefont {T.~J.}\ \bibnamefont {Kippenberg}},\ }\href@noop {}
  {\bibfield  {journal} {\bibinfo  {journal} {Nat. Commun.}\ }\textbf {\bibinfo
  {volume} {11}},\ \bibinfo {pages} {2402} (\bibinfo {year}
  {2020})}\BibitemShut {NoStop}%
\bibitem [{\citenamefont {Neshev}\ \emph {et~al.}(2003)\citenamefont {Neshev},
  \citenamefont {Ostrovskaya}, \citenamefont {Kivshar},\ and\ \citenamefont
  {Krolikowski}}]{Neshev:03}%
  \BibitemOpen
  \bibfield  {author} {\bibinfo {author} {\bibfnamefont {D.}~\bibnamefont
  {Neshev}}, \bibinfo {author} {\bibfnamefont {E.}~\bibnamefont {Ostrovskaya}},
  \bibinfo {author} {\bibfnamefont {Y.}~\bibnamefont {Kivshar}}, \ and\
  \bibinfo {author} {\bibfnamefont {W.}~\bibnamefont {Krolikowski}},\
  }\href@noop {} {\bibfield  {journal} {\bibinfo  {journal} {Opt. Lett.}\
  }\textbf {\bibinfo {volume} {28}},\ \bibinfo {pages} {710} (\bibinfo {year}
  {2003})}\BibitemShut {NoStop}%
\bibitem [{\citenamefont {Afanasjev}\ and\ \citenamefont
  {Vysloukh}(1994)}]{Afanasjev:94}%
  \BibitemOpen
  \bibfield  {author} {\bibinfo {author} {\bibfnamefont {V.~V.}\ \bibnamefont
  {Afanasjev}}\ and\ \bibinfo {author} {\bibfnamefont {V.~A.}\ \bibnamefont
  {Vysloukh}},\ }\href@noop {} {\bibfield  {journal} {\bibinfo  {journal} {J.
  Opt. Soc. Am. B}\ }\textbf {\bibinfo {volume} {11}},\ \bibinfo {pages} {2385}
  (\bibinfo {year} {1994})}\BibitemShut {NoStop}%
\bibitem [{\citenamefont {Tsoy}\ and\ \citenamefont
  {Abdullaev}(2003)}]{PhysRevE.67.056610}%
  \BibitemOpen
  \bibfield  {author} {\bibinfo {author} {\bibfnamefont {E.~N.}\ \bibnamefont
  {Tsoy}}\ and\ \bibinfo {author} {\bibfnamefont {F.~K.}\ \bibnamefont
  {Abdullaev}},\ }\href@noop {} {\bibfield  {journal} {\bibinfo  {journal}
  {Phys. Rev. E}\ }\textbf {\bibinfo {volume} {67}},\ \bibinfo {pages} {056610}
  (\bibinfo {year} {2003})}\BibitemShut {NoStop}%
\bibitem [{\citenamefont {Chen}\ \emph {et~al.}(2000)\citenamefont {Chen},
  \citenamefont {Acks}, \citenamefont {Ostrovskaya},\ and\ \citenamefont
  {Kivshar}}]{Chen:00}%
  \BibitemOpen
  \bibfield  {author} {\bibinfo {author} {\bibfnamefont {Z.}~\bibnamefont
  {Chen}}, \bibinfo {author} {\bibfnamefont {M.}~\bibnamefont {Acks}}, \bibinfo
  {author} {\bibfnamefont {E.~A.}\ \bibnamefont {Ostrovskaya}}, \ and\ \bibinfo
  {author} {\bibfnamefont {Y.~S.}\ \bibnamefont {Kivshar}},\ }\href@noop {}
  {\bibfield  {journal} {\bibinfo  {journal} {Opt. Lett.}\ }\textbf {\bibinfo
  {volume} {25}},\ \bibinfo {pages} {417} (\bibinfo {year} {2000})}\BibitemShut
  {NoStop}%
\bibitem [{\citenamefont {Kartashov}\ \emph {et~al.}(2002)\citenamefont
  {Kartashov}, \citenamefont {Crasovan}, \citenamefont {Mihalache},\ and\
  \citenamefont {Torner}}]{PhysRevLett.89.273902}%
  \BibitemOpen
  \bibfield  {author} {\bibinfo {author} {\bibfnamefont {Y.~V.}\ \bibnamefont
  {Kartashov}}, \bibinfo {author} {\bibfnamefont {L.-C.}\ \bibnamefont
  {Crasovan}}, \bibinfo {author} {\bibfnamefont {D.}~\bibnamefont {Mihalache}},
  \ and\ \bibinfo {author} {\bibfnamefont {L.}~\bibnamefont {Torner}},\
  }\href@noop {} {\bibfield  {journal} {\bibinfo  {journal} {Phys. Rev. Lett.}\
  }\textbf {\bibinfo {volume} {89}},\ \bibinfo {pages} {273902} (\bibinfo
  {year} {2002})}\BibitemShut {NoStop}%
\bibitem [{\citenamefont {Xu}\ \emph {et~al.}(2019)\citenamefont {Xu},
  \citenamefont {Gelash}, \citenamefont {Chabchoub}, \citenamefont {Zakharov},\
  and\ \citenamefont {Kibler}}]{PhysRevLett.122.084101}%
  \BibitemOpen
  \bibfield  {author} {\bibinfo {author} {\bibfnamefont {G.}~\bibnamefont
  {Xu}}, \bibinfo {author} {\bibfnamefont {A.}~\bibnamefont {Gelash}}, \bibinfo
  {author} {\bibfnamefont {A.}~\bibnamefont {Chabchoub}}, \bibinfo {author}
  {\bibfnamefont {V.}~\bibnamefont {Zakharov}}, \ and\ \bibinfo {author}
  {\bibfnamefont {B.}~\bibnamefont {Kibler}},\ }\href@noop {} {\bibfield
  {journal} {\bibinfo  {journal} {Phys. Rev. Lett.}\ }\textbf {\bibinfo
  {volume} {122}},\ \bibinfo {pages} {084101} (\bibinfo {year}
  {2019})}\BibitemShut {NoStop}%
\bibitem [{\citenamefont {Liu}\ \emph {et~al.}(2018)\citenamefont {Liu},
  \citenamefont {Yao},\ and\ \citenamefont {Cui}}]{PhysRevLett.121.023905}%
  \BibitemOpen
  \bibfield  {author} {\bibinfo {author} {\bibfnamefont {X.}~\bibnamefont
  {Liu}}, \bibinfo {author} {\bibfnamefont {X.}~\bibnamefont {Yao}}, \ and\
  \bibinfo {author} {\bibfnamefont {Y.}~\bibnamefont {Cui}},\ }\href@noop {}
  {\bibfield  {journal} {\bibinfo  {journal} {Phys. Rev. Lett.}\ }\textbf
  {\bibinfo {volume} {121}},\ \bibinfo {pages} {023905} (\bibinfo {year}
  {2018})}\BibitemShut {NoStop}%
\bibitem [{\citenamefont {Krupa}\ \emph {et~al.}(2017)\citenamefont {Krupa},
  \citenamefont {Nithyanandan}, \citenamefont {Andral}, \citenamefont
  {Tchofo-Dinda},\ and\ \citenamefont {Grelu}}]{PhysRevLett.118.243901}%
  \BibitemOpen
  \bibfield  {author} {\bibinfo {author} {\bibfnamefont {K.}~\bibnamefont
  {Krupa}}, \bibinfo {author} {\bibfnamefont {K.}~\bibnamefont {Nithyanandan}},
  \bibinfo {author} {\bibfnamefont {U.}~\bibnamefont {Andral}}, \bibinfo
  {author} {\bibfnamefont {P.}~\bibnamefont {Tchofo-Dinda}}, \ and\ \bibinfo
  {author} {\bibfnamefont {P.}~\bibnamefont {Grelu}},\ }\href@noop {}
  {\bibfield  {journal} {\bibinfo  {journal} {Phys. Rev. Lett.}\ }\textbf
  {\bibinfo {volume} {118}},\ \bibinfo {pages} {243901} (\bibinfo {year}
  {2017})}\BibitemShut {NoStop}%
\bibitem [{\citenamefont {Al~Khawaja}\ and\ \citenamefont
  {Boudjem\^aa}(2012)}]{PhysRevE.86.036606}%
  \BibitemOpen
  \bibfield  {author} {\bibinfo {author} {\bibfnamefont {U.}~\bibnamefont
  {Al~Khawaja}}\ and\ \bibinfo {author} {\bibfnamefont {A.}~\bibnamefont
  {Boudjem\^aa}},\ }\href@noop {} {\bibfield  {journal} {\bibinfo  {journal}
  {Phys. Rev. E}\ }\textbf {\bibinfo {volume} {86}},\ \bibinfo {pages} {036606}
  (\bibinfo {year} {2012})}\BibitemShut {NoStop}%
\bibitem [{\citenamefont {Grelu}\ and\ \citenamefont
  {Akhmediev}(2012)}]{Grelu_NatPhoton}%
  \BibitemOpen
  \bibfield  {author} {\bibinfo {author} {\bibfnamefont {P.}~\bibnamefont
  {Grelu}}\ and\ \bibinfo {author} {\bibfnamefont {N.}~\bibnamefont
  {Akhmediev}},\ }\href@noop {} {\bibfield  {journal} {\bibinfo  {journal}
  {Nat. Photonics}\ }\textbf {\bibinfo {volume} {6}},\ \bibinfo {pages} {84}
  (\bibinfo {year} {2012})}\BibitemShut {NoStop}%
\bibitem [{\citenamefont {Turmanov}\ \emph {et~al.}(2015)\citenamefont
  {Turmanov}, \citenamefont {Baizakov}, \citenamefont {Umarov},\ and\
  \citenamefont {Abdullaev}}]{TURMANOV20151828}%
  \BibitemOpen
  \bibfield  {author} {\bibinfo {author} {\bibfnamefont {B.}~\bibnamefont
  {Turmanov}}, \bibinfo {author} {\bibfnamefont {B.}~\bibnamefont {Baizakov}},
  \bibinfo {author} {\bibfnamefont {B.}~\bibnamefont {Umarov}}, \ and\ \bibinfo
  {author} {\bibfnamefont {F.}~\bibnamefont {Abdullaev}},\ }\href@noop {}
  {\bibfield  {journal} {\bibinfo  {journal} {Phys. Lett. A}\ }\textbf
  {\bibinfo {volume} {379}},\ \bibinfo {pages} {1828} (\bibinfo {year}
  {2015})}\BibitemShut {NoStop}%
\bibitem [{\citenamefont {Melchert}\ \emph {et~al.}(2019)\citenamefont
  {Melchert}, \citenamefont {Willms}, \citenamefont {Bose}, \citenamefont
  {Yulin}, \citenamefont {Roth}, \citenamefont {Mitschke}, \citenamefont
  {Morgner}, \citenamefont {Babushkin},\ and\ \citenamefont
  {Demircan}}]{PhysRevLett.123.243905}%
  \BibitemOpen
  \bibfield  {author} {\bibinfo {author} {\bibfnamefont {O.}~\bibnamefont
  {Melchert}}, \bibinfo {author} {\bibfnamefont {S.}~\bibnamefont {Willms}},
  \bibinfo {author} {\bibfnamefont {S.}~\bibnamefont {Bose}}, \bibinfo {author}
  {\bibfnamefont {A.}~\bibnamefont {Yulin}}, \bibinfo {author} {\bibfnamefont
  {B.}~\bibnamefont {Roth}}, \bibinfo {author} {\bibfnamefont {F.}~\bibnamefont
  {Mitschke}}, \bibinfo {author} {\bibfnamefont {U.}~\bibnamefont {Morgner}},
  \bibinfo {author} {\bibfnamefont {I.}~\bibnamefont {Babushkin}}, \ and\
  \bibinfo {author} {\bibfnamefont {A.}~\bibnamefont {Demircan}},\ }\href@noop
  {} {\bibfield  {journal} {\bibinfo  {journal} {Phys. Rev. Lett.}\ }\textbf
  {\bibinfo {volume} {123}},\ \bibinfo {pages} {243905} (\bibinfo {year}
  {2019})}\BibitemShut {NoStop}%
\bibitem [{\citenamefont {Tam}\ \emph {et~al.}(2020)\citenamefont {Tam},
  \citenamefont {Alexander}, \citenamefont {Blanco-Redondo},\ and\
  \citenamefont {de~Sterke}}]{PhysRevA.101.043822}%
  \BibitemOpen
  \bibfield  {author} {\bibinfo {author} {\bibfnamefont {K.~K.~K.}\
  \bibnamefont {Tam}}, \bibinfo {author} {\bibfnamefont {T.~J.}\ \bibnamefont
  {Alexander}}, \bibinfo {author} {\bibfnamefont {A.}~\bibnamefont
  {Blanco-Redondo}}, \ and\ \bibinfo {author} {\bibfnamefont {C.~M.}\
  \bibnamefont {de~Sterke}},\ }\href@noop {} {\bibfield  {journal} {\bibinfo
  {journal} {Phys. Rev. A}\ }\textbf {\bibinfo {volume} {101}},\ \bibinfo
  {pages} {043822} (\bibinfo {year} {2020})}\BibitemShut {NoStop}%
\bibitem [{\citenamefont {Melchert}\ and\ \citenamefont
  {Demircan}(2021)}]{Melchert:21}%
  \BibitemOpen
  \bibfield  {author} {\bibinfo {author} {\bibfnamefont {O.}~\bibnamefont
  {Melchert}}\ and\ \bibinfo {author} {\bibfnamefont {A.}~\bibnamefont
  {Demircan}},\ }\href@noop {} {\bibfield  {journal} {\bibinfo  {journal} {Opt.
  Lett.}\ }\textbf {\bibinfo {volume} {46}},\ \bibinfo {pages} {5603} (\bibinfo
  {year} {2021})}\BibitemShut {NoStop}%
\bibitem [{\citenamefont {Melchert}\ \emph {et~al.}(2021)\citenamefont
  {Melchert}, \citenamefont {Willms}, \citenamefont {Morgner}, \citenamefont
  {Babushkin},\ and\ \citenamefont {Demircan}}]{Melchert_SciRep21}%
  \BibitemOpen
  \bibfield  {author} {\bibinfo {author} {\bibfnamefont {O.}~\bibnamefont
  {Melchert}}, \bibinfo {author} {\bibfnamefont {S.}~\bibnamefont {Willms}},
  \bibinfo {author} {\bibfnamefont {U.}~\bibnamefont {Morgner}}, \bibinfo
  {author} {\bibfnamefont {I.}~\bibnamefont {Babushkin}}, \ and\ \bibinfo
  {author} {\bibfnamefont {A.}~\bibnamefont {Demircan}},\ }\href@noop {}
  {\bibfield  {journal} {\bibinfo  {journal} {Sci. Rep.}\ }\textbf {\bibinfo
  {volume} {11}},\ \bibinfo {pages} {11190} (\bibinfo {year}
  {2021})}\BibitemShut {NoStop}%
\bibitem [{\citenamefont {Lourdesamy}\ \emph {et~al.}(2022)\citenamefont
  {Lourdesamy}, \citenamefont {Runge}, \citenamefont {Alexander}, \citenamefont
  {Hudson}, \citenamefont {Blanco-Redondo},\ and\ \citenamefont
  {de~Sterke}}]{Lourdesamy_NaturePhys}%
  \BibitemOpen
  \bibfield  {author} {\bibinfo {author} {\bibfnamefont {J.~P.}\ \bibnamefont
  {Lourdesamy}}, \bibinfo {author} {\bibfnamefont {A.~F.~J.}\ \bibnamefont
  {Runge}}, \bibinfo {author} {\bibfnamefont {T.~J.}\ \bibnamefont
  {Alexander}}, \bibinfo {author} {\bibfnamefont {D.~D.}\ \bibnamefont
  {Hudson}}, \bibinfo {author} {\bibfnamefont {A.}~\bibnamefont
  {Blanco-Redondo}}, \ and\ \bibinfo {author} {\bibfnamefont {C.~M.}\
  \bibnamefont {de~Sterke}},\ }\href@noop {} {\bibfield  {journal} {\bibinfo
  {journal} {Nat. Phys.}\ }\textbf {\bibinfo {volume} {18}},\ \bibinfo {pages}
  {59} (\bibinfo {year} {2022})}\BibitemShut {NoStop}%
\bibitem [{\citenamefont {Moille}\ \emph {et~al.}(2018)\citenamefont {Moille},
  \citenamefont {Li}, \citenamefont {Kim}, \citenamefont {Westly},\ and\
  \citenamefont {Srinivasan}}]{Moille:18}%
  \BibitemOpen
  \bibfield  {author} {\bibinfo {author} {\bibfnamefont {G.}~\bibnamefont
  {Moille}}, \bibinfo {author} {\bibfnamefont {Q.}~\bibnamefont {Li}}, \bibinfo
  {author} {\bibfnamefont {S.}~\bibnamefont {Kim}}, \bibinfo {author}
  {\bibfnamefont {D.}~\bibnamefont {Westly}}, \ and\ \bibinfo {author}
  {\bibfnamefont {K.}~\bibnamefont {Srinivasan}},\ }\href@noop {} {\bibfield
  {journal} {\bibinfo  {journal} {Opt. Lett.}\ }\textbf {\bibinfo {volume}
  {43}},\ \bibinfo {pages} {2772} (\bibinfo {year} {2018})}\BibitemShut
  {NoStop}%
\bibitem [{\citenamefont {Melchert}\ \emph {et~al.}(2020)\citenamefont
  {Melchert}, \citenamefont {Yulin},\ and\ \citenamefont
  {Demircan}}]{Melchert:20}%
  \BibitemOpen
  \bibfield  {author} {\bibinfo {author} {\bibfnamefont {O.}~\bibnamefont
  {Melchert}}, \bibinfo {author} {\bibfnamefont {A.}~\bibnamefont {Yulin}}, \
  and\ \bibinfo {author} {\bibfnamefont {A.}~\bibnamefont {Demircan}},\
  }\href@noop {} {\bibfield  {journal} {\bibinfo  {journal} {Opt. Lett.}\
  }\textbf {\bibinfo {volume} {45}},\ \bibinfo {pages} {2764} (\bibinfo {year}
  {2020})}\BibitemShut {NoStop}%
\bibitem [{\citenamefont {Yulin}\ \emph {et~al.}(2004)\citenamefont {Yulin},
  \citenamefont {Skryabin},\ and\ \citenamefont {Russell}}]{Yulin:04}%
  \BibitemOpen
  \bibfield  {author} {\bibinfo {author} {\bibfnamefont {A.~V.}\ \bibnamefont
  {Yulin}}, \bibinfo {author} {\bibfnamefont {D.~V.}\ \bibnamefont {Skryabin}},
  \ and\ \bibinfo {author} {\bibfnamefont {P.~S.~J.}\ \bibnamefont {Russell}},\
  }\href@noop {} {\bibfield  {journal} {\bibinfo  {journal} {Opt. Lett.}\
  }\textbf {\bibinfo {volume} {29}},\ \bibinfo {pages} {2411} (\bibinfo {year}
  {2004})}\BibitemShut {NoStop}%
\bibitem [{\citenamefont {Skryabin}\ and\ \citenamefont
  {Yulin}(2005)}]{PhysRevE.72.016619}%
  \BibitemOpen
  \bibfield  {author} {\bibinfo {author} {\bibfnamefont {D.~V.}\ \bibnamefont
  {Skryabin}}\ and\ \bibinfo {author} {\bibfnamefont {A.~V.}\ \bibnamefont
  {Yulin}},\ }\href@noop {} {\bibfield  {journal} {\bibinfo  {journal} {Phys.
  Rev. E}\ }\textbf {\bibinfo {volume} {72}},\ \bibinfo {pages} {016619}
  (\bibinfo {year} {2005})}\BibitemShut {NoStop}%
\bibitem [{\citenamefont {Efimov}\ \emph {et~al.}(2004)\citenamefont {Efimov},
  \citenamefont {Taylor}, \citenamefont {Omenetto}, \citenamefont {Yulin},
  \citenamefont {Joly}, \citenamefont {Biancalana}, \citenamefont {Skryabin},
  \citenamefont {Knight},\ and\ \citenamefont {Russell}}]{Efimov:04}%
  \BibitemOpen
  \bibfield  {author} {\bibinfo {author} {\bibfnamefont {A.}~\bibnamefont
  {Efimov}}, \bibinfo {author} {\bibfnamefont {A.~J.}\ \bibnamefont {Taylor}},
  \bibinfo {author} {\bibfnamefont {F.~G.}\ \bibnamefont {Omenetto}}, \bibinfo
  {author} {\bibfnamefont {A.~V.}\ \bibnamefont {Yulin}}, \bibinfo {author}
  {\bibfnamefont {N.~Y.}\ \bibnamefont {Joly}}, \bibinfo {author}
  {\bibfnamefont {F.}~\bibnamefont {Biancalana}}, \bibinfo {author}
  {\bibfnamefont {D.~V.}\ \bibnamefont {Skryabin}}, \bibinfo {author}
  {\bibfnamefont {J.~C.}\ \bibnamefont {Knight}}, \ and\ \bibinfo {author}
  {\bibfnamefont {P.~S.}\ \bibnamefont {Russell}},\ }\href@noop {} {\bibfield
  {journal} {\bibinfo  {journal} {Opt. Express}\ }\textbf {\bibinfo {volume}
  {12}},\ \bibinfo {pages} {6498} (\bibinfo {year} {2004})}\BibitemShut
  {NoStop}%
\bibitem [{\citenamefont {Efimov}\ \emph {et~al.}(2005)\citenamefont {Efimov},
  \citenamefont {Yulin}, \citenamefont {Skryabin}, \citenamefont {Knight},
  \citenamefont {Joly}, \citenamefont {Omenetto}, \citenamefont {Taylor},\ and\
  \citenamefont {Russell}}]{PhysRevLett.95.213902}%
  \BibitemOpen
  \bibfield  {author} {\bibinfo {author} {\bibfnamefont {A.}~\bibnamefont
  {Efimov}}, \bibinfo {author} {\bibfnamefont {A.~V.}\ \bibnamefont {Yulin}},
  \bibinfo {author} {\bibfnamefont {D.~V.}\ \bibnamefont {Skryabin}}, \bibinfo
  {author} {\bibfnamefont {J.~C.}\ \bibnamefont {Knight}}, \bibinfo {author}
  {\bibfnamefont {N.}~\bibnamefont {Joly}}, \bibinfo {author} {\bibfnamefont
  {F.~G.}\ \bibnamefont {Omenetto}}, \bibinfo {author} {\bibfnamefont {A.~J.}\
  \bibnamefont {Taylor}}, \ and\ \bibinfo {author} {\bibfnamefont
  {P.}~\bibnamefont {Russell}},\ }\href@noop {} {\bibfield  {journal} {\bibinfo
   {journal} {Phys. Rev. Lett.}\ }\textbf {\bibinfo {volume} {95}},\ \bibinfo
  {pages} {213902} (\bibinfo {year} {2005})}\BibitemShut {NoStop}%
\bibitem [{\citenamefont {Efimov}\ \emph {et~al.}(2006)\citenamefont {Efimov},
  \citenamefont {Taylor}, \citenamefont {Yulin}, \citenamefont {Skryabin},\
  and\ \citenamefont {Knight}}]{Efimov:06}%
  \BibitemOpen
  \bibfield  {author} {\bibinfo {author} {\bibfnamefont {A.}~\bibnamefont
  {Efimov}}, \bibinfo {author} {\bibfnamefont {A.~J.}\ \bibnamefont {Taylor}},
  \bibinfo {author} {\bibfnamefont {A.~V.}\ \bibnamefont {Yulin}}, \bibinfo
  {author} {\bibfnamefont {D.~V.}\ \bibnamefont {Skryabin}}, \ and\ \bibinfo
  {author} {\bibfnamefont {J.~C.}\ \bibnamefont {Knight}},\ }\href@noop {}
  {\bibfield  {journal} {\bibinfo  {journal} {Opt. Lett.}\ }\textbf {\bibinfo
  {volume} {31}},\ \bibinfo {pages} {1624} (\bibinfo {year}
  {2006})}\BibitemShut {NoStop}%
\bibitem [{\citenamefont {de~Sterke}(1992)}]{deSterke:92}%
  \BibitemOpen
  \bibfield  {author} {\bibinfo {author} {\bibfnamefont {C.~M.}\ \bibnamefont
  {de~Sterke}},\ }\href@noop {} {\bibfield  {journal} {\bibinfo  {journal}
  {Opt. Lett.}\ }\textbf {\bibinfo {volume} {17}},\ \bibinfo {pages} {914}
  (\bibinfo {year} {1992})}\BibitemShut {NoStop}%
\bibitem [{\citenamefont {Gorbach}\ and\ \citenamefont
  {Skryabin}(2007)}]{Gorbach_NatPhoton}%
  \BibitemOpen
  \bibfield  {author} {\bibinfo {author} {\bibfnamefont {A.~V.}\ \bibnamefont
  {Gorbach}}\ and\ \bibinfo {author} {\bibfnamefont {D.~V.}\ \bibnamefont
  {Skryabin}},\ }\href@noop {} {\bibfield  {journal} {\bibinfo  {journal} {Nat.
  Photonics}\ }\textbf {\bibinfo {volume} {1}},\ \bibinfo {pages} {653}
  (\bibinfo {year} {2007})}\BibitemShut {NoStop}%
\bibitem [{\citenamefont {Demircan}\ \emph {et~al.}(2013)\citenamefont
  {Demircan}, \citenamefont {Amiranashvili}, \citenamefont {Br\'ee},\ and\
  \citenamefont {Steinmeyer}}]{PhysRevLett.110.233901}%
  \BibitemOpen
  \bibfield  {author} {\bibinfo {author} {\bibfnamefont {A.}~\bibnamefont
  {Demircan}}, \bibinfo {author} {\bibfnamefont {S.}~\bibnamefont
  {Amiranashvili}}, \bibinfo {author} {\bibfnamefont {C.}~\bibnamefont
  {Br\'ee}}, \ and\ \bibinfo {author} {\bibfnamefont {G.}~\bibnamefont
  {Steinmeyer}},\ }\href@noop {} {\bibfield  {journal} {\bibinfo  {journal}
  {Phys. Rev. Lett.}\ }\textbf {\bibinfo {volume} {110}},\ \bibinfo {pages}
  {233901} (\bibinfo {year} {2013})}\BibitemShut {NoStop}%
\bibitem [{\citenamefont {Demircan}\ \emph {et~al.}(2012)\citenamefont
  {Demircan}, \citenamefont {Amiranashvili}, \citenamefont {Br\'ee},
  \citenamefont {Mahnke}, \citenamefont {Mitschke},\ and\ \citenamefont
  {Steinmeyer}}]{Demircan_SciRep}%
  \BibitemOpen
  \bibfield  {author} {\bibinfo {author} {\bibfnamefont {A.}~\bibnamefont
  {Demircan}}, \bibinfo {author} {\bibfnamefont {S.}~\bibnamefont
  {Amiranashvili}}, \bibinfo {author} {\bibfnamefont {C.}~\bibnamefont
  {Br\'ee}}, \bibinfo {author} {\bibfnamefont {C.}~\bibnamefont {Mahnke}},
  \bibinfo {author} {\bibfnamefont {F.}~\bibnamefont {Mitschke}}, \ and\
  \bibinfo {author} {\bibfnamefont {G.}~\bibnamefont {Steinmeyer}},\
  }\href@noop {} {\bibfield  {journal} {\bibinfo  {journal} {Sci. Rep.}\
  }\textbf {\bibinfo {volume} {2}},\ \bibinfo {pages} {850} (\bibinfo {year}
  {2012})}\BibitemShut {NoStop}%
\bibitem [{\citenamefont {Philbin}\ \emph {et~al.}(2008)\citenamefont
  {Philbin}, \citenamefont {Kuklewicz}, \citenamefont {Robertson},
  \citenamefont {Hill}, \citenamefont {Konig},\ and\ \citenamefont
  {Leonhardt}}]{Philbin_Science}%
  \BibitemOpen
  \bibfield  {author} {\bibinfo {author} {\bibfnamefont {T.~G.}\ \bibnamefont
  {Philbin}}, \bibinfo {author} {\bibfnamefont {C.}~\bibnamefont {Kuklewicz}},
  \bibinfo {author} {\bibfnamefont {S.}~\bibnamefont {Robertson}}, \bibinfo
  {author} {\bibfnamefont {S.}~\bibnamefont {Hill}}, \bibinfo {author}
  {\bibfnamefont {F.}~\bibnamefont {Konig}}, \ and\ \bibinfo {author}
  {\bibfnamefont {U.}~\bibnamefont {Leonhardt}},\ }\href@noop {} {\bibfield
  {journal} {\bibinfo  {journal} {Science}\ }\textbf {\bibinfo {volume}
  {319}},\ \bibinfo {pages} {1367} (\bibinfo {year} {2008})}\BibitemShut
  {NoStop}%
\bibitem [{\citenamefont {Demircan}\ \emph {et~al.}(2011)\citenamefont
  {Demircan}, \citenamefont {Amiranashvili},\ and\ \citenamefont
  {Steinmeyer}}]{PhysRevLett.106.163901}%
  \BibitemOpen
  \bibfield  {author} {\bibinfo {author} {\bibfnamefont {A.}~\bibnamefont
  {Demircan}}, \bibinfo {author} {\bibfnamefont {S.}~\bibnamefont
  {Amiranashvili}}, \ and\ \bibinfo {author} {\bibfnamefont {G.}~\bibnamefont
  {Steinmeyer}},\ }\href@noop {} {\bibfield  {journal} {\bibinfo  {journal}
  {Phys. Rev. Lett.}\ }\textbf {\bibinfo {volume} {106}},\ \bibinfo {pages}
  {163901} (\bibinfo {year} {2011})}\BibitemShut {NoStop}%
\bibitem [{\citenamefont {Yulin}\ \emph {et~al.}(2013)\citenamefont {Yulin},
  \citenamefont {Driben}, \citenamefont {Malomed},\ and\ \citenamefont
  {Skryabin}}]{Yulin:13}%
  \BibitemOpen
  \bibfield  {author} {\bibinfo {author} {\bibfnamefont {A.~V.}\ \bibnamefont
  {Yulin}}, \bibinfo {author} {\bibfnamefont {R.}~\bibnamefont {Driben}},
  \bibinfo {author} {\bibfnamefont {B.~A.}\ \bibnamefont {Malomed}}, \ and\
  \bibinfo {author} {\bibfnamefont {D.~V.}\ \bibnamefont {Skryabin}},\
  }\href@noop {} {\bibfield  {journal} {\bibinfo  {journal} {Opt. Express}\
  }\textbf {\bibinfo {volume} {21}},\ \bibinfo {pages} {14481} (\bibinfo {year}
  {2013})}\BibitemShut {NoStop}%
\bibitem [{\citenamefont {Demircan}\ \emph {et~al.}(2014)\citenamefont
  {Demircan}, \citenamefont {Amiranashvili}, \citenamefont {Br\'{e}e},
  \citenamefont {Morgner},\ and\ \citenamefont {Steinmeyer}}]{Demircan:14}%
  \BibitemOpen
  \bibfield  {author} {\bibinfo {author} {\bibfnamefont {A.}~\bibnamefont
  {Demircan}}, \bibinfo {author} {\bibfnamefont {S.}~\bibnamefont
  {Amiranashvili}}, \bibinfo {author} {\bibfnamefont {C.}~\bibnamefont
  {Br\'{e}e}}, \bibinfo {author} {\bibfnamefont {U.}~\bibnamefont {Morgner}}, \
  and\ \bibinfo {author} {\bibfnamefont {G.}~\bibnamefont {Steinmeyer}},\
  }\href@noop {} {\bibfield  {journal} {\bibinfo  {journal} {Opt. Lett.}\
  }\textbf {\bibinfo {volume} {39}},\ \bibinfo {pages} {2735} (\bibinfo {year}
  {2014})}\BibitemShut {NoStop}%
\bibitem [{\citenamefont {Wang}\ \emph {et~al.}(2015)\citenamefont {Wang},
  \citenamefont {Mussot}, \citenamefont {Conforti}, \citenamefont {Zeng},\ and\
  \citenamefont {Kudlinski}}]{Wang:15}%
  \BibitemOpen
  \bibfield  {author} {\bibinfo {author} {\bibfnamefont {S.~F.}\ \bibnamefont
  {Wang}}, \bibinfo {author} {\bibfnamefont {A.}~\bibnamefont {Mussot}},
  \bibinfo {author} {\bibfnamefont {M.}~\bibnamefont {Conforti}}, \bibinfo
  {author} {\bibfnamefont {X.~L.}\ \bibnamefont {Zeng}}, \ and\ \bibinfo
  {author} {\bibfnamefont {A.}~\bibnamefont {Kudlinski}},\ }\href@noop {}
  {\bibfield  {journal} {\bibinfo  {journal} {Opt. Lett.}\ }\textbf {\bibinfo
  {volume} {40}},\ \bibinfo {pages} {3320} (\bibinfo {year}
  {2015})}\BibitemShut {NoStop}%
\bibitem [{\citenamefont {Yulin}(2018)}]{PhysRevA.98.023833}%
  \BibitemOpen
  \bibfield  {author} {\bibinfo {author} {\bibfnamefont {A.~V.}\ \bibnamefont
  {Yulin}},\ }\href@noop {} {\bibfield  {journal} {\bibinfo  {journal} {Phys.
  Rev. A}\ }\textbf {\bibinfo {volume} {98}},\ \bibinfo {pages} {023833}
  (\bibinfo {year} {2018})}\BibitemShut {NoStop}%
\bibitem [{\citenamefont {Oreshnikov}\ \emph
  {et~al.}(2015{\natexlab{a}})\citenamefont {Oreshnikov}, \citenamefont
  {Driben},\ and\ \citenamefont {Yulin}}]{Oreshnikov:15}%
  \BibitemOpen
  \bibfield  {author} {\bibinfo {author} {\bibfnamefont {I.}~\bibnamefont
  {Oreshnikov}}, \bibinfo {author} {\bibfnamefont {R.}~\bibnamefont {Driben}},
  \ and\ \bibinfo {author} {\bibfnamefont {A.~V.}\ \bibnamefont {Yulin}},\
  }\href@noop {} {\bibfield  {journal} {\bibinfo  {journal} {Opt. Lett.}\
  }\textbf {\bibinfo {volume} {40}},\ \bibinfo {pages} {4871} (\bibinfo {year}
  {2015}{\natexlab{a}})}\BibitemShut {NoStop}%
\bibitem [{\citenamefont {Marest}\ \emph {et~al.}(2018)\citenamefont {Marest},
  \citenamefont {Arab\'{i}}, \citenamefont {Conforti}, \citenamefont {Mussot},
  \citenamefont {Mili\'{a}n}, \citenamefont {Skryabin},\ and\ \citenamefont
  {Kudlinski}}]{Marest:18}%
  \BibitemOpen
  \bibfield  {author} {\bibinfo {author} {\bibfnamefont {T.}~\bibnamefont
  {Marest}}, \bibinfo {author} {\bibfnamefont {C.~M.}\ \bibnamefont
  {Arab\'{i}}}, \bibinfo {author} {\bibfnamefont {M.}~\bibnamefont {Conforti}},
  \bibinfo {author} {\bibfnamefont {A.}~\bibnamefont {Mussot}}, \bibinfo
  {author} {\bibfnamefont {C.}~\bibnamefont {Mili\'{a}n}}, \bibinfo {author}
  {\bibfnamefont {D.~V.}\ \bibnamefont {Skryabin}}, \ and\ \bibinfo {author}
  {\bibfnamefont {A.}~\bibnamefont {Kudlinski}},\ }\href@noop {} {\bibfield
  {journal} {\bibinfo  {journal} {Opt. Express}\ }\textbf {\bibinfo {volume}
  {26}},\ \bibinfo {pages} {23480} (\bibinfo {year} {2018})}\BibitemShut
  {NoStop}%
\bibitem [{\citenamefont {Deng}\ \emph {et~al.}(2018)\citenamefont {Deng},
  \citenamefont {Liu}, \citenamefont {Huang}, \citenamefont {Zhao},\ and\
  \citenamefont {Wang}}]{Deng:18}%
  \BibitemOpen
  \bibfield  {author} {\bibinfo {author} {\bibfnamefont {Z.}~\bibnamefont
  {Deng}}, \bibinfo {author} {\bibfnamefont {J.}~\bibnamefont {Liu}}, \bibinfo
  {author} {\bibfnamefont {X.}~\bibnamefont {Huang}}, \bibinfo {author}
  {\bibfnamefont {C.}~\bibnamefont {Zhao}}, \ and\ \bibinfo {author}
  {\bibfnamefont {X.}~\bibnamefont {Wang}},\ }\href@noop {} {\bibfield
  {journal} {\bibinfo  {journal} {Opt. Express}\ }\textbf {\bibinfo {volume}
  {26}},\ \bibinfo {pages} {16535} (\bibinfo {year} {2018})}\BibitemShut
  {NoStop}%
\bibitem [{\citenamefont {Rong}\ \emph {et~al.}(2021)\citenamefont {Rong},
  \citenamefont {Yang}, \citenamefont {Xiao},\ and\ \citenamefont
  {Chen}}]{PhysRevA.103.023505}%
  \BibitemOpen
  \bibfield  {author} {\bibinfo {author} {\bibfnamefont {J.}~\bibnamefont
  {Rong}}, \bibinfo {author} {\bibfnamefont {H.}~\bibnamefont {Yang}}, \bibinfo
  {author} {\bibfnamefont {Y.}~\bibnamefont {Xiao}}, \ and\ \bibinfo {author}
  {\bibfnamefont {Y.}~\bibnamefont {Chen}},\ }\href@noop {} {\bibfield
  {journal} {\bibinfo  {journal} {Phys. Rev. A}\ }\textbf {\bibinfo {volume}
  {103}},\ \bibinfo {pages} {023505} (\bibinfo {year} {2021})}\BibitemShut
  {NoStop}%
\bibitem [{\citenamefont {Kodama}\ \emph {et~al.}(1994)\citenamefont {Kodama},
  \citenamefont {Romagnoli}, \citenamefont {Wabnitz},\ and\ \citenamefont
  {Midrio}}]{Kodama:94}%
  \BibitemOpen
  \bibfield  {author} {\bibinfo {author} {\bibfnamefont {Y.}~\bibnamefont
  {Kodama}}, \bibinfo {author} {\bibfnamefont {M.}~\bibnamefont {Romagnoli}},
  \bibinfo {author} {\bibfnamefont {S.}~\bibnamefont {Wabnitz}}, \ and\
  \bibinfo {author} {\bibfnamefont {M.}~\bibnamefont {Midrio}},\ }\href@noop {}
  {\bibfield  {journal} {\bibinfo  {journal} {Opt. Lett.}\ }\textbf {\bibinfo
  {volume} {19}},\ \bibinfo {pages} {165} (\bibinfo {year} {1994})}\BibitemShut
  {NoStop}%
\bibitem [{\citenamefont {Conforti}\ \emph {et~al.}(2015)\citenamefont
  {Conforti}, \citenamefont {Trillo}, \citenamefont {Mussot},\ and\
  \citenamefont {Kudlinski}}]{Conforti_SciRep}%
  \BibitemOpen
  \bibfield  {author} {\bibinfo {author} {\bibfnamefont {M.}~\bibnamefont
  {Conforti}}, \bibinfo {author} {\bibfnamefont {S.}~\bibnamefont {Trillo}},
  \bibinfo {author} {\bibfnamefont {A.}~\bibnamefont {Mussot}}, \ and\ \bibinfo
  {author} {\bibfnamefont {A.}~\bibnamefont {Kudlinski}},\ }\href@noop {}
  {\bibfield  {journal} {\bibinfo  {journal} {Sci. Rep.}\ }\textbf {\bibinfo
  {volume} {5}},\ \bibinfo {pages} {9433} (\bibinfo {year} {2015})}\BibitemShut
  {NoStop}%
\bibitem [{\citenamefont {Driben}\ \emph {et~al.}(2015)\citenamefont {Driben},
  \citenamefont {Yulin},\ and\ \citenamefont {Efimov}}]{Driben:15}%
  \BibitemOpen
  \bibfield  {author} {\bibinfo {author} {\bibfnamefont {R.}~\bibnamefont
  {Driben}}, \bibinfo {author} {\bibfnamefont {A.~V.}\ \bibnamefont {Yulin}}, \
  and\ \bibinfo {author} {\bibfnamefont {A.}~\bibnamefont {Efimov}},\
  }\href@noop {} {\bibfield  {journal} {\bibinfo  {journal} {Opt. Express}\
  }\textbf {\bibinfo {volume} {23}},\ \bibinfo {pages} {19112} (\bibinfo {year}
  {2015})}\BibitemShut {NoStop}%
\bibitem [{\citenamefont {Wright}\ \emph {et~al.}(2015)\citenamefont {Wright},
  \citenamefont {Wabnitz}, \citenamefont {Christodoulides},\ and\ \citenamefont
  {Wise}}]{PhysRevLett.115.223902}%
  \BibitemOpen
  \bibfield  {author} {\bibinfo {author} {\bibfnamefont {L.~G.}\ \bibnamefont
  {Wright}}, \bibinfo {author} {\bibfnamefont {S.}~\bibnamefont {Wabnitz}},
  \bibinfo {author} {\bibfnamefont {D.~N.}\ \bibnamefont {Christodoulides}}, \
  and\ \bibinfo {author} {\bibfnamefont {F.~W.}\ \bibnamefont {Wise}},\
  }\href@noop {} {\bibfield  {journal} {\bibinfo  {journal} {Phys. Rev. Lett.}\
  }\textbf {\bibinfo {volume} {115}},\ \bibinfo {pages} {223902} (\bibinfo
  {year} {2015})}\BibitemShut {NoStop}%
\bibitem [{\citenamefont {Krupa}\ \emph {et~al.}(2016)\citenamefont {Krupa},
  \citenamefont {Tonello}, \citenamefont {Barth\'el\'emy}, \citenamefont
  {Couderc}, \citenamefont {Shalaby}, \citenamefont {Bendahmane}, \citenamefont
  {Millot},\ and\ \citenamefont {Wabnitz}}]{PhysRevLett.116.183901}%
  \BibitemOpen
  \bibfield  {author} {\bibinfo {author} {\bibfnamefont {K.}~\bibnamefont
  {Krupa}}, \bibinfo {author} {\bibfnamefont {A.}~\bibnamefont {Tonello}},
  \bibinfo {author} {\bibfnamefont {A.}~\bibnamefont {Barth\'el\'emy}},
  \bibinfo {author} {\bibfnamefont {V.}~\bibnamefont {Couderc}}, \bibinfo
  {author} {\bibfnamefont {B.~M.}\ \bibnamefont {Shalaby}}, \bibinfo {author}
  {\bibfnamefont {A.}~\bibnamefont {Bendahmane}}, \bibinfo {author}
  {\bibfnamefont {G.}~\bibnamefont {Millot}}, \ and\ \bibinfo {author}
  {\bibfnamefont {S.}~\bibnamefont {Wabnitz}},\ }\href@noop {} {\bibfield
  {journal} {\bibinfo  {journal} {Phys. Rev. Lett.}\ }\textbf {\bibinfo
  {volume} {116}},\ \bibinfo {pages} {183901} (\bibinfo {year}
  {2016})}\BibitemShut {NoStop}%
\bibitem [{\citenamefont {Oreshnikov}\ \emph {et~al.}(2017)\citenamefont
  {Oreshnikov}, \citenamefont {Driben},\ and\ \citenamefont
  {Yulin}}]{PhysRevA.96.013809}%
  \BibitemOpen
  \bibfield  {author} {\bibinfo {author} {\bibfnamefont {I.}~\bibnamefont
  {Oreshnikov}}, \bibinfo {author} {\bibfnamefont {R.}~\bibnamefont {Driben}},
  \ and\ \bibinfo {author} {\bibfnamefont {A.}~\bibnamefont {Yulin}},\
  }\href@noop {} {\bibfield  {journal} {\bibinfo  {journal} {Phys. Rev. A}\
  }\textbf {\bibinfo {volume} {96}},\ \bibinfo {pages} {013809} (\bibinfo
  {year} {2017})}\BibitemShut {NoStop}%
\bibitem [{\citenamefont {Akhmediev}\ and\ \citenamefont
  {Karlsson}(1995)}]{akhmediev1995cherenkov}%
  \BibitemOpen
  \bibfield  {author} {\bibinfo {author} {\bibfnamefont {N.}~\bibnamefont
  {Akhmediev}}\ and\ \bibinfo {author} {\bibfnamefont {M.}~\bibnamefont
  {Karlsson}},\ }\href@noop {} {\bibfield  {journal} {\bibinfo  {journal}
  {Phys. Rev. A}\ }\textbf {\bibinfo {volume} {51}},\ \bibinfo {pages} {2602}
  (\bibinfo {year} {1995})}\BibitemShut {NoStop}%
\bibitem [{\citenamefont {Afanasjev}\ \emph {et~al.}(1996)\citenamefont
  {Afanasjev}, \citenamefont {Kivshar},\ and\ \citenamefont
  {Menyuk}}]{afanasjev1996effect}%
  \BibitemOpen
  \bibfield  {author} {\bibinfo {author} {\bibfnamefont {V.~V.}\ \bibnamefont
  {Afanasjev}}, \bibinfo {author} {\bibfnamefont {Y.~S.}\ \bibnamefont
  {Kivshar}}, \ and\ \bibinfo {author} {\bibfnamefont {C.~R.}\ \bibnamefont
  {Menyuk}},\ }\href@noop {} {\bibfield  {journal} {\bibinfo  {journal} {Opt.
  Lett.}\ }\textbf {\bibinfo {volume} {21}},\ \bibinfo {pages} {1975} (\bibinfo
  {year} {1996})}\BibitemShut {NoStop}%
\bibitem [{\citenamefont {Amiranashvili}\ and\ \citenamefont
  {Demircan}(2010)}]{amiranashvili2010hamiltonian}%
  \BibitemOpen
  \bibfield  {author} {\bibinfo {author} {\bibfnamefont {S.}~\bibnamefont
  {Amiranashvili}}\ and\ \bibinfo {author} {\bibfnamefont {A.}~\bibnamefont
  {Demircan}},\ }\href@noop {} {\bibfield  {journal} {\bibinfo  {journal}
  {Physical Review A}\ }\textbf {\bibinfo {volume} {82}},\ \bibinfo {pages}
  {013812} (\bibinfo {year} {2010})}\BibitemShut {NoStop}%
\bibitem [{\citenamefont {Agrawal}(2013)}]{agrawal2013nonlinear}%
  \BibitemOpen
  \bibfield  {author} {\bibinfo {author} {\bibfnamefont {G.~P.}\ \bibnamefont
  {Agrawal}}\ }(\bibinfo  {publisher} {Academic Press},\ \bibinfo {year}
  {2013})\ p.\ \bibinfo {pages} {247}\BibitemShut {NoStop}%
\bibitem [{\citenamefont {Hindmarsch}(1983)}]{hindmarsh1983odepack}%
  \BibitemOpen
  \bibfield  {author} {\bibinfo {author} {\bibfnamefont {A.~C.}\ \bibnamefont
  {Hindmarsch}},\ }in\ \href@noop {} {\emph {\bibinfo {booktitle} {Scientific
  Computing}}},\ \bibinfo {editor} {edited by\ \bibinfo {editor} {\bibfnamefont
  {R.~S.}\ \bibnamefont {Stepleman}}, \bibinfo {editor} {\bibfnamefont
  {M.}~\bibnamefont {Carver}}, \bibinfo {editor} {\bibfnamefont
  {R.}~\bibnamefont {Peskin}}, \bibinfo {editor} {\bibfnamefont {W.~F.}\
  \bibnamefont {Ames}}, \ and\ \bibinfo {editor} {\bibfnamefont
  {R.}~\bibnamefont {Vichnevetsky}}}\ (\bibinfo  {publisher} {North-Holland
  Publishing Company},\ \bibinfo {address} {Amsterdam},\ \bibinfo {year}
  {1983})\ pp.\ \bibinfo {pages} {55--64}\BibitemShut {NoStop}%
\bibitem [{\citenamefont {Virtanen}\ \emph {et~al.}(2020)\citenamefont
  {Virtanen}, \citenamefont {Gommers}, \citenamefont {Oliphant}, \citenamefont
  {Haberland}, \citenamefont {Reddy}, \citenamefont {Cournapeau}, \citenamefont
  {Burovski}, \citenamefont {Peterson}, \citenamefont {Weckesser},
  \citenamefont {Bright} \emph {et~al.}}]{virtanen2020scipy}%
  \BibitemOpen
  \bibfield  {author} {\bibinfo {author} {\bibfnamefont {P.}~\bibnamefont
  {Virtanen}}, \bibinfo {author} {\bibfnamefont {R.}~\bibnamefont {Gommers}},
  \bibinfo {author} {\bibfnamefont {T.~E.}\ \bibnamefont {Oliphant}}, \bibinfo
  {author} {\bibfnamefont {M.}~\bibnamefont {Haberland}}, \bibinfo {author}
  {\bibfnamefont {T.}~\bibnamefont {Reddy}}, \bibinfo {author} {\bibfnamefont
  {D.}~\bibnamefont {Cournapeau}}, \bibinfo {author} {\bibfnamefont
  {E.}~\bibnamefont {Burovski}}, \bibinfo {author} {\bibfnamefont
  {P.}~\bibnamefont {Peterson}}, \bibinfo {author} {\bibfnamefont
  {W.}~\bibnamefont {Weckesser}}, \bibinfo {author} {\bibfnamefont
  {J.}~\bibnamefont {Bright}},  \emph {et~al.},\ }\href@noop {} {\bibfield
  {journal} {\bibinfo  {journal} {Nature methods}\ }\textbf {\bibinfo {volume}
  {17}},\ \bibinfo {pages} {261} (\bibinfo {year} {2020})}\BibitemShut
  {NoStop}%
\bibitem [{sou()}]{sources}%
  \BibitemOpen
  \href@noop {} {}\bibinfo {howpublished}
  {\url{https://github.com/ioreshnikov/two-color-solitons}}\BibitemShut
  {NoStop}%
\bibitem [{sup()}]{suppMat}%
  \BibitemOpen
  \href@noop {} {\ }\bibinfo {note} {See Supplemental Material at [URL will be
  inserted by publisher] for a derivation of the resonance wavenumber of the
  two-color solitons internal oscillation mode.}\BibitemShut {Stop}%
\bibitem [{\citenamefont {Tartara}(2015)}]{tartara2015soliton}%
  \BibitemOpen
  \bibfield  {author} {\bibinfo {author} {\bibfnamefont {L.}~\bibnamefont
  {Tartara}},\ }\href@noop {} {\bibfield  {journal} {\bibinfo  {journal} {JOSA
  B}\ }\textbf {\bibinfo {volume} {32}},\ \bibinfo {pages} {395} (\bibinfo
  {year} {2015})}\BibitemShut {NoStop}%
\bibitem [{\citenamefont {Oreshnikov}\ \emph
  {et~al.}(2015{\natexlab{b}})\citenamefont {Oreshnikov}, \citenamefont
  {Driben},\ and\ \citenamefont {Yulin}}]{oreshnikov2015interaction}%
  \BibitemOpen
  \bibfield  {author} {\bibinfo {author} {\bibfnamefont {I.}~\bibnamefont
  {Oreshnikov}}, \bibinfo {author} {\bibfnamefont {R.}~\bibnamefont {Driben}},
  \ and\ \bibinfo {author} {\bibfnamefont {A.}~\bibnamefont {Yulin}},\
  }\href@noop {} {\bibfield  {journal} {\bibinfo  {journal} {Opt. Lett.}\
  }\textbf {\bibinfo {volume} {40}},\ \bibinfo {pages} {5554} (\bibinfo {year}
  {2015}{\natexlab{b}})}\BibitemShut {NoStop}%
\bibitem [{\citenamefont {Willms}\ \emph {et~al.}(2022)\citenamefont {Willms},
  \citenamefont {Melchert}, \citenamefont {Bose}, \citenamefont {Yulin},
  \citenamefont {Oreshnikov}, \citenamefont {Morgner}, \citenamefont
  {Babushkin},\ and\ \citenamefont {Demircan}}]{Willms:PRA:2022}%
  \BibitemOpen
  \bibfield  {author} {\bibinfo {author} {\bibfnamefont {S.}~\bibnamefont
  {Willms}}, \bibinfo {author} {\bibfnamefont {O.}~\bibnamefont {Melchert}},
  \bibinfo {author} {\bibfnamefont {S.}~\bibnamefont {Bose}}, \bibinfo {author}
  {\bibfnamefont {A.}~\bibnamefont {Yulin}}, \bibinfo {author} {\bibfnamefont
  {I.}~\bibnamefont {Oreshnikov}}, \bibinfo {author} {\bibfnamefont
  {U.}~\bibnamefont {Morgner}}, \bibinfo {author} {\bibfnamefont
  {I.}~\bibnamefont {Babushkin}}, \ and\ \bibinfo {author} {\bibfnamefont
  {A.}~\bibnamefont {Demircan}},\ }\href@noop {} {\bibfield  {journal}
  {\bibinfo  {journal} {Phys. Rev. A}\ }\textbf {\bibinfo {volume} {105}},\
  \bibinfo {pages} {053525} (\bibinfo {year} {2022})}\BibitemShut {NoStop}%
\end{thebibliography}%

\end{document}


\title{Cherenkov radiation and scattering of external dispersive waves by two-color solitons (Supplemental Material)}

\author{Ivan Oreshnikov}
\affiliation{Max Planck Institute for Intelligent Systems, Max-Planck-Ring 4, 72076 T\"ubingen, Germany}

\author{Oliver Melchert}
\affiliation{Cluster of Exellence PhoenixD, Welfengarten 1, 30167 Hannover, Germany}
\affiliation{Leibniz Universit\"at Hannover, Institute of Quantum Optics, Welfengarten 1, 30167 Hannover, Germany}
\affiliation{Hannover Centre for Optical Technologies, Nienburger Strasse 17, 30167 Hannover, Germany}

\author{Stephanie Willms}
\affiliation{Cluster of Exellence PhoenixD, Welfengarten 1, 30167 Hannover, Germany}
\affiliation{Leibniz Universit\"at Hannover, Institute of Quantum Optics, Welfengarten 1, 30167 Hannover, Germany}

\author{Surajit Bose}
\affiliation{Cluster of Exellence PhoenixD, Welfengarten 1, 30167 Hannover, Germany}
\affiliation{Leibniz Universit\"at Hannover, Institute of Photonics, Nienburger Strasse 17, 30167 Hannover, Germany}

\author{Ihar Babushkin}
\affiliation{Cluster of Exellence PhoenixD, Welfengarten 1, 30167 Hannover, Germany}
\affiliation{Leibniz Universit\"at Hannover, Institute of Quantum Optics, Welfengarten 1, 30167 Hannover, Germany}

\author{Ayhan Demircan}
\affiliation{Cluster of Exellence PhoenixD, Welfengarten 1, 30167 Hannover, Germany}
\affiliation{Leibniz Universit\"at Hannover, Institute of Quantum Optics, Welfengarten 1, 30167 Hannover, Germany}
\affiliation{Hannover Centre for Optical Technologies, Nienburger Strasse 17, 30167 Hannover, Germany}

\author{Uwe Morgner}
\affiliation{Cluster of Exellence PhoenixD, Welfengarten 1, 30167 Hannover, Germany}
\affiliation{Leibniz Universit\"at Hannover, Institute of Quantum Optics, Welfengarten 1, 30167 Hannover, Germany}
\affiliation{Hannover Centre for Optical Technologies, Nienburger Strasse 17, 30167 Hannover, Germany}

\author{Alexey Yulin}
\affiliation{Department of Physics, ITMO University, Kronverskiy pr. 49, 19701 St.~Petersburg, Russia}

\date{\today}

\maketitle

\section*{Dispersion profile model}

\begin{figure}[ht]
  \centering
  \includegraphics{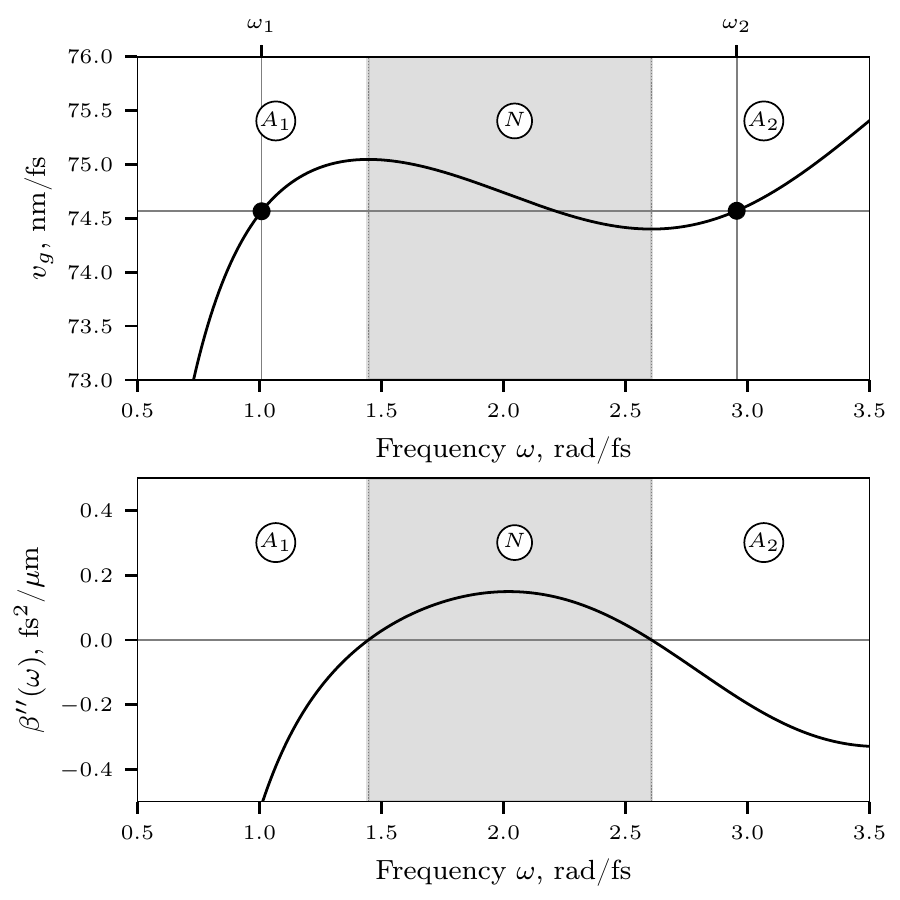}
  \caption{
    (a) group velocity $v_{g}$ and (b) second order dispersion coefficient $\beta''(\omega)$ in the model fiber. Labels $A_{1,2}$ and $N$ mark the regions of anomalous and normal dispersion.
  }
  \label{fig:Model}
\end{figure}

We model dispersion coefficient $\beta(\omega)$ with the following rational expression
\begin{equation}
  \label{eq:Beta}
  \beta(\omega) = \frac{1}{c} \frac{
    \sum_{n=0}^{3} {C_{n} \omega^{n+1}}
  }{
    \sum_{m=0}^{3} {D_{m} \omega^{m}}
    }
\end{equation}
where $c = 0.299792458~\mu
\text{m/fs}$ is the speed of light, and the coefficient sequences $C$ and $D$ are
defined by
\begin{eqnarray}
C &=& (9.654, -39.739 \,\mathrm{fs},  16.885 \,\mathrm{fs^2}, -2.746 \,\mathrm{fs^3}),\quad \text{and},\\
D &=& (1, -9.496 \,\mathrm{fs}, 4.221 \,\mathrm{fs^2}, -0.703 \,\mathrm{fs^3}).
\end{eqnarray}
In here and everywhere in the paper we assume fs as a unit of
time and $\mu$m as a unit of distance. Figure \ref{fig:Model} displays group
velocity $v_{g}(\omega) = 1 / \beta'(\omega)$ and second order dispersion
coefficient $\beta''(\omega)$ as functions of frequency. Frequencies
$\omega_{1}$ and $\omega_{2}$ correspond to the central frequencies of the
soliton's spectral components as chosen in the simulation corresponding to
Fig.\,1 of the paper.

\section*{Small internal oscillations of the soliton}

When analyzing the nonlinear scattering near an oscillatory mode we used an expression for the oscillation frequency. In this section we will derive this expression.

Let us return to Eq.~(3) (of the main text) for coupled solitons. We can re-normalize the equations by performing the following transformation
\begin{equation*}
  U_{n} \to \gamma_{n}^{1/2} e^{i \beta_{n} z} \cdot U_{n},
\end{equation*}
which will make the equations symmetric
\begin{equation*}
  i \partial_{z} U_{n}
    - \frac{1}{2} \beta''_{n} \partial_{t}^{2} U_{n}
    + \gamma_{n}^{2} \abs{U_{n}}^{2} U_{n}
    + 2 \gamma_{n} \gamma_{m} \abs{U_{m}}^{2} U_{n} = 0.
\end{equation*}
This in turn allows us to recognize the modified couple of equations as Euler-Lagrange equations for Lagrangian
\begin{equation*}
  \int \limits_{-\infty}^{+\infty}
    \mathcal{L}(U_{1}, \partial_{z} U_{1}, \partial_{t} U_{1}, \ldots)
    \, dt,
\end{equation*}
where Lagrangian density $\mathcal{L}$ is defined as a sum of three components $\mathcal{L} = \mathcal{L}_{1} + \mathcal{L}_{2} + \mathcal{L}_{\text{int}}$, with $\mathcal{L}_{n}$ being a single-soliton Lagrangian density
\begin{equation}
  \label{eq:SingleSolitonLagrangianDensity}
  \mathcal{L}_{n} =
  \frac{i}{2} \left(
    \partial_{z} U_{n} \cdot U_{n}^{*} -
    \partial_{z} U_{n}^{*} \cdot U_{n}
  \right)
  + \frac{1}{2} \beta''_{n} \, \partial_{t} U_{n} \partial_{t} U_{n}^{*}
  + \frac{1}{2} \gamma_{n}^{2} \, \abs{U_{n}}^{4},
\end{equation}
and $\mathcal{L}_{\text{int}}$ being the interaction term
\begin{equation}
  \label{eq:InteractionLagrangianDensity}
  \mathcal{L}_{\text{int}} =
    2 \gamma_{1} \gamma_{2}
    \abs{U_{1}}^{2} \abs{U_{2}}^{2}.
\end{equation}
Let us assume that the soliton components $U_{n}$ can be described by the following generic \textit{ansatz}
\begin{equation}
  \label{eq:SolitonAnsatz}
  U_{n}(z, t) = A_{n}(z) S\left(
    \frac{t - t_{n}(z)}{\sigma_{n}(z)}
  \right) \exp\left(
    - i \Omega_{n}(z) t + i \phi_{n}(z)
  \right).
\end{equation}
In here $A_{n}$ is the amplitude of the pulse, $t_{n}$ is the central position, $\sigma_{n}$ is the pulse width, $\Omega_{n}$ is the frequency detuning, $\phi_{n}$ is the phase, and $S(x)$ is function that defines the envelope shape. At the moment we will not specify the concrete form of $S(x)$, but will assume that it is an even function.

Before we continue let us stress one important thing: this ansatz cannot express all the possible internal oscillations of the soliton. One obvious example, as it was noted in the text, is the case of the pulse-width oscillation. In order to capture this dynamics, we need to add frequency chirp to the ansatz.

Substituting \eqref{eq:SolitonAnsatz} into \eqref{eq:SingleSolitonLagrangianDensity} and \eqref{eq:InteractionLagrangianDensity} and integrating over $t$ we arrive at the expressions for the averaged Lagrangians
\begin{gather}
  L_{n} =
    I_{1} \, t_{n} \sigma_{n} A_{n}^{2} \frac{d \Omega_{n}}{d z} +
    I_{1} \, \sigma_{n} A_{n}^{2} \frac{d \phi_{n}}{d z} +
    I_{2} \, \frac{\beta''_{n}}{2} \frac{A_{n}^{2}}{\sigma_{n}} \nonumber \\
    + I_{1} \, \frac{\beta''_{n}}{2} \sigma_{n} A_{n}^{2} \Omega_{n}^{2} +
    I_{3} \, \frac{\gamma_{n}^{2}}{2} \sigma_{n} A_{n}^{4} \\
  L_{\text{int}} =
    2 \, \gamma_{1} \gamma_{2} \, A_{1}^{2} A_{2}^{2} \,
    I_{\text{int}}(\sigma_{1}, \sigma_{2}, t_{1}, t_{2}),
\end{gather}
where the following integrals have been defined
\begin{align*}
  I_{1} =
    \int \limits_{-\infty}^{+\infty}
    S^{2}(x) \, dx &&
  I_{2} =
    \int \limits_{-\infty}^{+\infty}
    (S'(x))^{2} \, dx \\
  I_{3} =
    \int \limits_{-\infty}^{+\infty}
    S^{4}(x) \, dx &&
  I_{\text{int}} =
    \int \limits_{-\infty}^{+\infty}
    S^{2}\left(
      \frac{t - t_{1}}{\sigma_{1}}
    \right)
    S^{2}\left(
      \frac{t - t_{2}}{\sigma_{2}}
    \right) \, dt
\end{align*}
Due to the time invariance in the problem, $I_{\text{int}}$ depends only on the difference between $t_{1}$ and $t_{2}$
\begin{equation*}
  I_{\text{int}} = I_{\text{int}}(t_{1} - t_{2}, \sigma_{1}, \sigma_{2}),
\end{equation*}
and it is an even function of that difference.

The averaged Lagrangian $L = L_{1} + L_{2} + L_{\text{int}}$ is now a function defined in terms of soliton parameters \{\,$A_{n}$, $\sigma_{n}$, $t_{n}$, $\Omega_{n}$, $\phi_{n}$\,\} and only them. Therefore, the Euler-Lagrange equations for the new Lagrangian have to be defined in terms of variations over the soliton parameters
\begin{equation*}
  \frac{\delta L}{\delta P_{n}} =
    \frac{\partial{L}}{{\partial P_{n}}} -
    \frac{d}{dz} \frac{\partial L}{\partial \dot P_{n}} = 0,
\end{equation*}
where $P_{n}$ stands for either $A_{n}$, $\sigma_{n}$, $t_{n}$, $\Omega_{n}$ or $\phi_{n}$. The latter case --- variation with respect to the phase $\phi_{n}$ --- immediately yields the conservation of mass
\begin{equation}
  \label{eq:ConservationOfMass}
  N_{n} = \sigma_{n}(z) A_{n}^{2}(z) = \mathrm{const.}
\end{equation}

Variation with respect to the detuning $\Omega_{n}$ fixes the group velocity of individual solitons
\begin{equation}
  \label{eq:SolitonPosition}
  \frac{d t_{n}}{dz} = \beta''_{n} \Omega_{n}(z).
\end{equation}

Variation with respect to the soliton position $t_{n}$ gives us an equation for the frequency
\begin{equation}
  \label{eq:SolitonFrequency}
  \frac{d \Omega_{n}}{dz} =
    2 \cdot \frac{N_{m} \gamma_{1} \gamma_{2}}
           {I_{1} \sigma_{1}(z) \sigma_{2}(z)} \cdot
    \frac{\partial I_{\text{int}}}{\partial t_{n}}.
\end{equation}
The symmetry in the overlap integral $I_{\text{int}}$ with respect to the soliton positions $t_{1}$ and $t_{2}$ leads to conservation of momentum
\begin{equation}
  \label{eq:ConservationOfMomentum}
  N_{1} \Omega_{1}(z) + N_{2} \Omega_{2}(z) = \mathrm{const}.
\end{equation}

Finally, the difference between the variations with respect to $A_{n}$ and $\sigma_{n}$ gives us
\begin{equation}
  \label{eq:SolitonWidth}
  I_{2} \beta''_{n} +
  \frac{I_{3} \gamma_{n}}{2} N_{n} \sigma_{n}(z)
  + 2 N_{m} \gamma_{1} \gamma_{2} \frac{\sigma_{n}(z)}{\sigma_{m}(z)} \left(
    I_{\text{int}} + \sigma_{n} \frac{\partial I_{\text{int}}}{\sigma_{n}}
  \right) = 0.
\end{equation}
The very last equation --- omitted here --- is the evolution equation for the phase $\phi_{n}$. The right-hand's side of the equation is quite complicated, but since the phase does not occur anywhere in \eqref{eq:SolitonPosition}, \eqref{eq:SolitonFrequency} or \eqref{eq:SolitonWidth}, it is not important for the remaining analysis.

Let us switch from the the individual soliton positions to the mean position and the relative delay instead
\begin{align*}
  t_{0} = \frac{1}{2} \left(
    t_{1} + t_{2}
  \right) &&
  \Delta t = t_{1} - t_{2}
\end{align*}
Equation for the relative delay $\Delta t$
\begin{equation}
  \label{eq:RelativeDelay}
  \frac{d \Delta t}{dz} =
    \beta''_{1} \Omega_{1}(z) +
    \beta''_{2} \Omega_{2}(z)
\end{equation}
and equations \eqref{eq:SolitonFrequency} and \eqref{eq:SolitonWidth} form a closed system, with equations for $d \Delta t / dz$, $d \Omega_{n} / dz$ acting as equations of motion and equations \eqref{eq:SolitonWidth} fixing the widths $\sigma_{n}(z)$ as functions of $\Delta t$. By differentiating \eqref{eq:RelativeDelay} one more time and using \eqref{eq:SolitonFrequency} we get
\begin{equation*}
  \frac{d^{2} \Delta t}{dz^{2}} +
  2 \frac{
    \gamma_{1} \gamma_{2} \left(
      \beta''_{1} N_{1} + \beta''_{2} N_{2}
    \right)
  }{
    I_{1} \sigma_{1}(\Delta t) \sigma_{2}(\Delta t)
  } \frac{\partial}{\partial \Delta t}
  I_{\text{int}} (\Delta t, \sigma_{1}, \sigma_{2}) = 0
\end{equation*}
To transform this into a harmonic oscillator equation we need to linearize the second term around the equilibrium point $\Delta t = 0$. Since $I_{\text{int}}$ is an even function, the derivative $\partial I_{\text{int}} / \partial \Delta {t}$ is odd and it vanishes at $\Delta t = 0$. This means we can ignore $\Delta t$ dependency in $\sigma_{1}$ and $\sigma_{2}$ --- only the term proportional to $\partial^{2} I_{\text{int}} / \partial \Delta t^{2}$ will survive. Thus we finally arrive at
\begin{equation*}
  \frac{d^{2} \Delta t}{dz^{2}} +
  K_{0}^{2} \Delta t = 0,
\end{equation*}
where the resonance frequency $K_{0}$ is
\begin{equation}
  \label{eq:ResonanceFrequency}
  K_{0}^{2} = 2 \frac{
    \gamma_{1} \gamma_{2} \left(
      \beta''_{1} N_{1} + \beta''_{2} N_{2}
    \right)
  }{
    I_{1} \sigma_{1}(0) \sigma_{2}(0)
  } I_{\text{int}}''(0; \sigma_{1}(0), \sigma_{2}(0)).
\end{equation}

For a more concrete estimate let us finally consider a Gaussian envelope, i.e.~let us set $S(x) = \exp(-x^{2})$. Such a choice of the envelope shape fixes the integrals $I_{1} = \sqrt{\pi / 2}$ and
\begin{equation*}
  I_{\text{int}}(\Delta t, \sigma_{1}, \sigma_{2}) =
  \sqrt{ \frac{\pi}{2} }
  \frac{
    \sigma_{1} \sigma_{2}
  }{
    \sqrt{
      \left(
        \sigma_{1}^{2} + \sigma_{2}^{2}
      \right)
    }
  }
  \cdot \exp \left(
    \frac{
      -2 \Delta t^{2}
    }{
      \sigma_{1}^{2} + \sigma_{2}^{2}
    }
  \right),
\end{equation*}
which finally gives us the following expression for the resonance frequency
\begin{equation}
  \label{eq:ResonanceFrequencyGaussian}
  K_{0}^{2} =
    - \frac{
      8 \, \gamma(\omega_{1}) \gamma(\omega_{2})
    }{
      \left(
        \sigma_{1}^{2} + \sigma_{2}^{2}
      \right)^{3/2}
    } \cdot \left(
      \beta''(\omega_{1}) \sigma_{1} A_{1}^{2} +
      \beta''(\omega_{2}) \sigma_{2} A_{2}^{2}
    \right).
\end{equation}